\documentclass{aa}
\usepackage[varg]{txfonts}
\usepackage{graphicx}
\usepackage{dblfloatfix}
\usepackage{epstopdf}
\usepackage{pdflscape}
\usepackage{threeparttable,booktabs}  
\usepackage{txfonts}
\usepackage{gensymb}
\usepackage{color}
\usepackage{url}
\usepackage{multirow}
\usepackage{amsmath}
\usepackage{amstext}
\usepackage{natbib}
\usepackage{longtable}
\usepackage{float}
\usepackage[export]{adjustbox}
\usepackage{caption}
\usepackage{siunitx} 
\usepackage{threeparttablex} 

\begin{document}

\title{Optical spectropolarimetry of large C-complex asteroids: polarimetric evidence for heterogeneous surface compositions}

\author{Yuna G. Kwon\inst{\ref{inst1},\ref{inst4}}~\and Stefano Bagnulo\inst{\ref{inst2}}~\and Alberto Cellino\inst{\ref{inst3}}
 }

\institute{Institut f{\" u}r Geophysik und Extraterrestrische Physik, Technische Universit{\" a}t Braunschweig,  Mendelssohnstr. 3, 38106 Braunschweig, Germany (\email{ynkwontop@gmail.com,y.kwon@tu-braunschweig.de}\label{inst1})
\and Armagh Observatory, College Hill, Armagh BT61 9DG, Northern Ireland, UK\label{inst2}
\and INAF -- Osservatorio Astrofisico di Torino, Pino Torinese, Italy\label{inst3}
\and Current affiliation: Caltech/IPAC, 1200 E California Blvd, MC 100-22, Pasadena, CA 91125, USA\label{inst4}
}

\date{Received May 26, 2023 / Accepted July 28, 2023}

\abstract {This study presents the first optical (420--930 nm) spectropolarimetric study of a sample of large ($\gtrsim$100 km in diameter) C-complex asteroids in observing circumstances for which light scattered by asteroid surfaces undergoes so-called negative polarization. A total of 64 C-complex asteroids of different subclasses are analyzed using archival polarimetric and reflectance data to refine the link between polarimetric parameters and surface properties of the asteroids.
We find a consistent difference in the polarization spectra between asteroids containing phyllosilicates and those without, which correlates with the overall morphology of the reflectance spectrum. The inverse relationship between reflectance and polarization degree, known as the Umov law, is not very sharp in our sample asteroids. They exhibit broad similarities in polarization-phase curves; nonetheless, we observe a gradual enhancement of the negative polarization branch (both in depth and width) in the ascending order of F $\rightarrow$ B $\rightarrow$ T $\rightarrow$ Ch types (Spearman correlation coefficient $\rho$ = 0.70$^{+0.06}_{-0.07}$ and p-value $p$ = 0.0001), along with an increase in reflectance curvature around 500 nm. Weaker inverse correlation between the inversion angle and diameter of the asteroids in our sample has been found with $\rho$ = $-$0.42$^{+0.08}_{-0.07}$ and $p$ = 0.01, apparently driven by the distribution of C-type asteroids ($\rho$ = $-$0.51$^{+0.13}_{-0.11}$ and $p$ = 0.04).
Our observations suggest that at least for large C-complex asteroids, 1) a common mechanism underlies the diversity in polarimetric and spectroscopic properties at optical wavelengths, 2) the observed trends would be explained by the surface composition of the asteroids, particularly optical heterogeneity on the wavelength scale caused by carbon's varying levels of optical influence, and 3) aqueous alteration may play a significant role in regulating this operational effect of carbon.}


\titlerunning{Optical spectropolarimetry of large C-complex asteroids by VLT/FORS2}

\authorrunning{Y. G. Kwon et al.}

\maketitle

\section{Introduction \label{sec:intro}}

As remnants of planet-forming processes, asteroids make up a negligible portion of the total mass of our solar system; nonetheless, their abundance and diversity in observations provide a fascinating glimpse into the intricacies of the early evolution of planetesimals. Among them, so-called C-complex asteroids with low albedo ($\lesssim$0.1; \citealt{Bowell1978,Tholen1984}) have been garnering steady attention due to their volatile-rich (icy) composition  \citep{Rivkin2015a} that might have an impact on habitability \citep{Kwok2018}. The presence of volatile materials (or traces of their past existence) makes C-complex asteroids useful for linking observations to their early history.

C-complex asteroids, which account for around half of the asteroid belt clustering between Mars and Jupiter \citep{DeMeo2014}, are classified into several taxonomic types \citep{Tholen1984,DeMeo2009} based on the properties of their reflectance spectra at visible and near-infrared wavelengths  (0.4--2.5 $\mu$m; VNIR). Variations in VNIR spectral properties can be attributed to both the inherent compositional and physical properties of grains on asteroidal surfaces (i.e., regoliths). Composition-wise, relatively featureless spectra with negative-to-moderately positive VNIR slopes make them historically connected with carbonaceous meteorites with varying contents of carbons, metals, and silicates \citep{Johnson1973}. Post-accretionary processes would alter the natal composition of the asteroids. In particular, aqueous alteration, a low-temperature process where melting ice interacts with anhydrous silicates and metals, has been observed across various subclasses of C-complex asteroids. This process produces unique absorption features, such as phyllosilicate bands near 0.7 and 3 $\mu$m (e.g. \citealt{Fornasier2014,Rivkin2022}), and/or changes spectral slopes in the VNIR \citep{Trang2021}. The physical makeup of regoliths, such as size and texture, is strongly dependent on the collisional history of the bodies whose characteristic lifetime is a function of the diameter \citep{Bottke2005}. While large ($\gtrsim$100 km) asteroids are regarded as primordial objects that have well-developed particulate regoliths over sufficient time, smaller ($<$100 km) asteroids are having undergone active surface rejuvenation and contain particles that are relatively larger in size due to the weak gravitational force of the bodies \citep{Dollfus1989,Delbo2007,Marchis2012}.

Polarimetry has long been used to study regolith environments. Asteroids scatter light in a manner that encodes details about their surfaces by interacting electromagnetically with the regolith particles. Upon scattering, the light becomes partially polarized, whose degree of polarization and its variations with phase angle $\alpha$ (angle of the Sun--asteroid--observer) and wavelength provide a useful diagnostics of the surface nature  \citep{Cellino2015a}. In particular, the reversal of polarization position angles around the backscattering region ($\alpha$ $\lesssim$ 30\degree) producing the negative polarization branch (NPB; \citealt{Shkuratov1985}) is one of the hallmarks of asteroids' regoliths. Instead of simple Fresnel's reflection, complicated interferences between multiple-scattered light by regolith particles (coherent backscattering mechanism; \citealt{Muinonen2015} and references therein) should be taken into account. Given that the shape of the NPB ties to compositional and textural information of regoliths, polarimetry has been successfully applied to lunar regolith studies \citep{Dollfus1975,Shkuratov1992,Dollfus1998}, asteroids' albedo estimation \citep{Zellner1977a,Geake1986,Cellino2015b}, and taxonomy studies \citep{Belskaya2017,Masiero2022}. 
Spectropolarimetry has emerged in recent years as a promising technique for exploring asteroid regoliths. \citet{Bagnulo2015} show that asteroids with similar reflectance spectra may exhibit significantly different polarization spectra, yet there have been no follow-up studies, primarily because polarimeters are more difficult to access than traditional imaging or spectroscopic facilities.

\begin{figure}[!h]
\centering
\includegraphics[width=8.5cm]{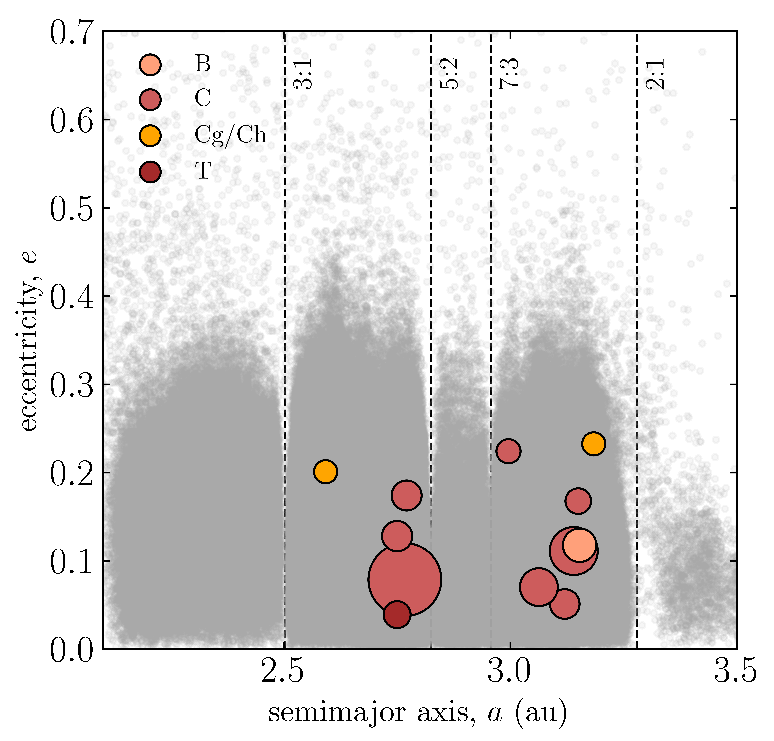}
\caption{Orbital distributions of 12 C-complex asteroids observed with VLT/FORS2 in spectropolarimetry Background grey dots are numbered main-belt asteroids that are quoted from JPL Small-Body Database Query (\protect\url{https://ssd.jpl.nasa.gov/tools/sbdb_query.html}). Symbols in the highlighted color are proportional to the size of asteroids. Vertical lines at different heliocentric distances indicate the most important mean-motion resonances with Jupiter.}
\label{Fig01}
\end{figure}

We here present new spectropolarimetric observations of large C-complex asteroids over 420--930 nm from the European Southern Observatory's Very Large Telescope (ESO/VLT) and the William Herschel Telescope (WHT). In this paper, we focus on large ($\gtrsim$100 km) low-albedo ($\lesssim$0.1) C-complex asteroids, including some that have not yet been fully explored from a polarimetric perspective. The adopted size criterion of the asteroids would help reduce the potential impact of different regolith sizes on optical behavior and secure strong enough signals to obtain accurate measurements of small amounts of polarization (typical maximum polarization in the NPB for asteroids is $\lesssim$2 \%; \citealt{Cellino2015a}).  We evaluate the spectropolarimetric properties of our 12 targets (whose orbital distributions are shown in Fig. \ref{Fig01}) and the correlation between the shapes of their polarization and reflectance spectra. With the aid of 52 additional C-complex asteroids acquired through polarimetric surveys, we examine the distribution of the inversion angle (defining the angular width of the NPB) and its relationship to the morphology of the reflectance spectra for a total of 64 asteroids belonging to different subclasses.

The restricted range in size and taxonomy of the objects considered in our analysis provides only a partial picture of the overall asteroid population. However, by presenting the first systematic polarimetric study of these low-albedo asteroid groups and showing how their polarimetric behavior presumably correlates with reflectance attributes, this study is intended to help ascribe the primary surface characteristics of asteroids influencing optical wavelength observations.

\section{Observations and data analysis \label{sec:obsdata}}

\begin{table*}[!t]
\centering
\caption{Journal of observations in spectropolarimetry}
\begin{tabular}{c|c|l|c|cc|cc}
\toprule
\hline
Telescope/ & \multirow{2}{*}{Mode} & \multirow{2}{*}{Ast. Name} & Median & $t$$^{\rm a}$   & $\alpha$$^{\rm b}$  & $r_{\rm H}$$^{\rm c}$  & $\Delta$$^{\rm d}$ \\
Instrument & & & UT Date \& Time & (sec) & ($^{\circ}$)  & (au) & (au) \\
\hline
\hline
\multirow{14}{*}{VLT-UT1/} & \multirow{14}{*}{PMOS} & (1) Ceres & 2016 09 12 09:18 & 64 & 14.5 & 2.903 & 2.115 \\
\cline{4-8}
\multirow{15}{*}{FORS2} & \multirow{15}{*}{GRIS\_300V} & (10) Hygiea & 2016 04 30 23:46 & 200 & 15.9 & 2.892 & 2.159 \\
\cline{4-8}
 & & (24) Themis & 2014 02 05 04:54 & 480 & 14.0 & 2.763 & 1.956 \\
\cline{4-8}
 & & \multirow{2}{*}{(35) Leukothea} & 2016 04 12 08:31 & 960 & 8.9 & 2.319 & 1.354 \\
 & & & 2016 07 06 02:01 & 1440 & 23.4 & 2.384 & 1.822 \\
 \cline{4-8}
 & & (90) Antiope & 2016 05 20 09:40 & 1040 & 8.8 & 2.822 & 1.873 \\
 \cline{4-8}
 & & (120) Lachesis & 2016 05 20 07:14 & 720 & 10.9 & 3.003 & 2.112 \\
 \cline{4-8}
 & & (128) Nemesis & 2016 05 15 01:00 & 720 & 9.0 & 3.083 & 2.158 \\
 \cline{4-8}
 & & (175) Andromache & 2015 06 02 05:04 & 2120 & 16.4 & 2.688 & 1.906 \\
 \cline{4-8}
 & & \multirow{2}{*}{(308) Polyxo} & 2015 05 24 09:15 & 1220 & 15.8 & 2.947 & 1.839 \\
 & & & 2015 09 08 01:20 & 1260 & 20.8 & 2.640 & 2.103 \\
 \cline{4-8}
 & & \multirow{2}{*}{(404) Arsinoe} & 2016 05 26 05:09 & 720 & 7.6 & 2.157 & 1.166 \\
 & & & 2016 08 07 01:47 & 1760 & 23.4 & 2.278 & 1.635 \\
 \cline{4-8}
 & & \multirow{2}{*}{(444) Gyptis} & 2015 05 22 03:24 & 1000 & 7.0 & 2.782 & 1.802 \\
 & & & 2015 08 07 01:39 & 800 & 21.2 & 2.641 & 2.117 \\
 \cline{1-8}
 \multirow{2}{*}{WHT/ISIS} & \multirow{2}{*}{R300B, R158R} & (1) Ceres & 2014 05 13 00:23 & 200 & 12.4 & 2.646 & 1.750 \\
  \cline{4-8}
 & & (451) Patientia & 2014 05 13 03:16 & 1200 & 12.4 & 3.220 & 2.416 \\
 \hline
 \hline
\end{tabular}
\tablefoot{$^{\rm a}$On-source integration time of each target in seconds; $^{\rm b}$Phase angle, in degree; $^{\rm c}$Heliocentric distance, in au; $^{\rm d}$Geocentric distance, in au. Each asteroid's observing geometry is provided by NASA/JPL Horizon (\url{https://ssd.jpl.nasa.gov/horizons/app.html#/}).} 
\label{t1}
\end{table*}

Spectropolarimetric observations of 12 C-complex asteroids were conducted over 2014--2016 using the FORS2 instrument mounted on the Cassegrain focus of the 8.0-m diameter UT1 telescope of the VLT at the Paranal Observatory in Chile. FORS2, FOcal Reducer and low-dispersion Spectrograph, is a versatile instrument that offers multi-object spectropolarimetry (PMOS) mode. The same instrumental setup was applied throughout all PMOS observations, using grism 300V with no order separating filter, and a 2\arcsec\ slit width, for a spectral coverage between 400 and 930 nm, and a resolving power of $\sim 220$. More details are given in \citep{Bagnulo2015}. 
Two asteroids were observed with the ISIS instrument of the WHT. Details about the instrument and observing technique are given in \citet{Bagnulo2019}. ISIS observations were obtained using grating R300B in the instrument blue arm (range 370--590 nm) and grating R158R plus order-sorting filter GG495 in the red arm (range 480–975 nm). A journal of observations is given in Table \ref{t1}.

\begin{figure*}[!t]
\centering
\includegraphics[width=\textwidth]{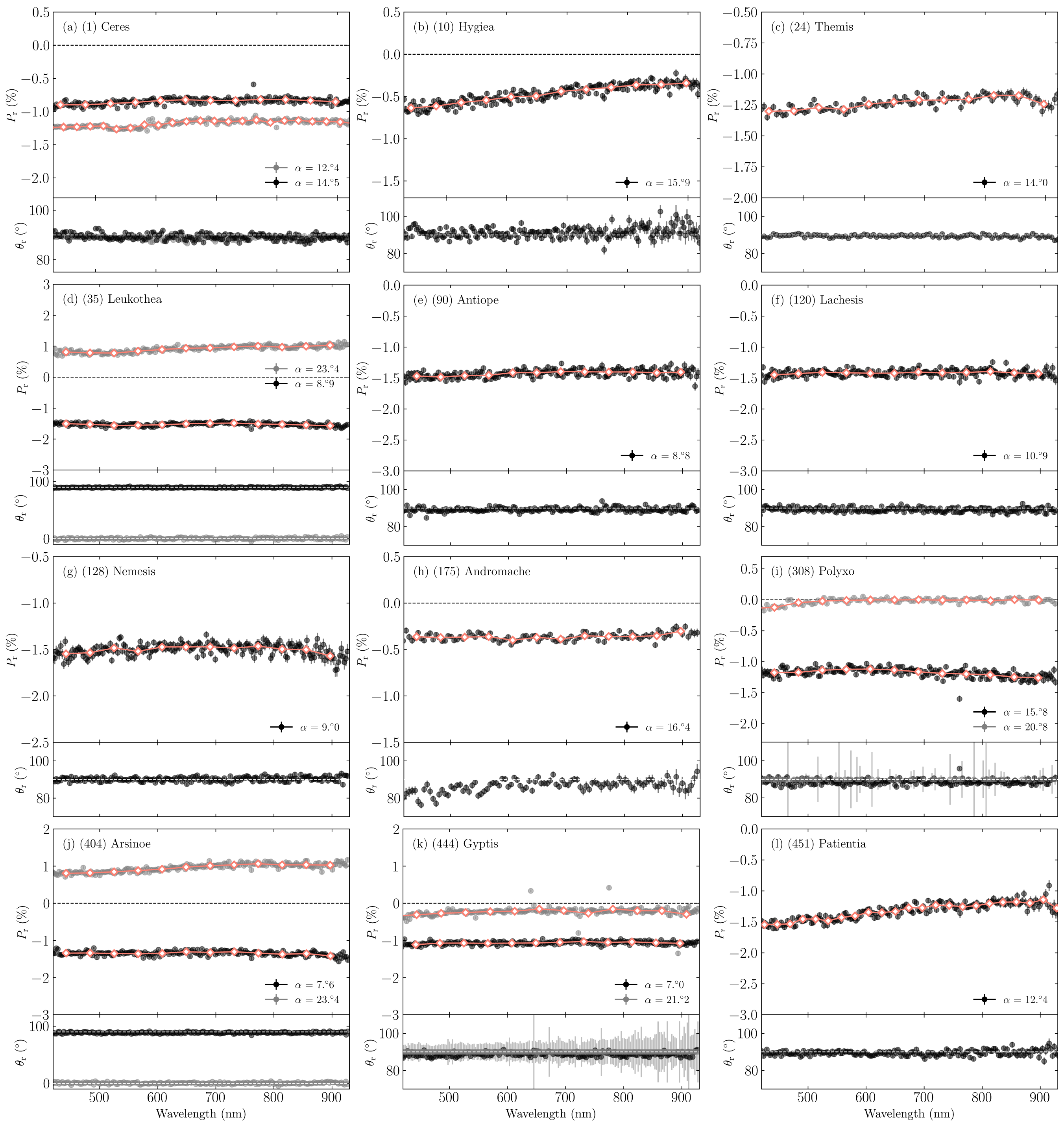}
\caption{Reduced polarization degree $P_{\rm r}$ and its position vector $\theta_{\rm r}$ as a function of wavelength for 12 C-complex asteroids taken at 17 epochs (Table \ref{t1}). In each panel, the phase angle ($\alpha$) at the time of observation is provided in the legend. Different colors were used for asteroids observed more than once. The curve with salmon-colored diamonds represents the binned polarization spectrum with the coarsest spectral resolution (an interval of $\sim$20--40 nm).}
\label{Fig02}
\end{figure*}

Raw images were preprocessed using standard methods (bias subtraction, cosmic-ray removal, and sky subtraction). The calibration and extraction of polarimetric parameters were conducted in the same manner as \citet{Bagnulo2015}: the observed Stokes profiles were rotated to take into account the chromatism of the retarder waveplate as explained, for instance, in Sect.~4.1 of \citet{Bagnulo2019}\footnote{For the VLT/FORS2 data, the instrumental angle information is provided by the FORS2 user manual and the webpage (https://www.eso.org/sci/facilities/paranal/instruments/fors/inst/pola.html).}. 
As is customary in solar system science, the corrected degree of linear polarization $P$ and its position angle $\theta_{\rm P}$ (with regard to the equatorial system; Eqs. 6 and 7 in \citealt{Kwon2022a}) were transformed into a system whose reference direction is perpendicular to the scattering plane (a plane including the Sun--asteroid--observer) using
\begin{equation}
P_{\rm r} = P~\cos(2\theta_{\rm r})
\label{eq:eq1}
\end{equation}
\noindent where
\begin{equation}
\theta_{\rm r} = \theta_{\rm P} - \Bigg(\phi \pm \frac{\pi}{2}\Bigg)
\label{eq:eq2}
\end{equation}
\noindent in the same manner as \citet{Chernova1993}. Here $\phi$ is the position angle of the scattering plane. Input $\phi$ was adjusted to satisfy 0 $\le$ ($\phi$ $\pm$ $\pi$/2) $\le$ $\pi$. The reduced (rotated) $P_{\rm r}$ value now has either a negative or positive sign depending on viewing geometry, that is, $P_{\rm r}$ $>$ 0 is the polarization perpendicular to the scattering plane and $P_{\rm r}$ $<$ 0 is the polarization parallel to the scattering plane. Accordingly, $\theta_{\rm r}$ should be ideally 0\degree\ in the positive polarization branch (PPB) and 90\degree\ in the negative polarization branch (NPB), which serves as an independent check for the reliability of our data reduction procedures. Polarization spectra $P_{\rm r}(\lambda)$ over 420--930 nm for our 12 C-complex asteroids are shown in Figure \ref{Fig02}, along with the spectral distribution of $\theta_{\rm r}$. 
All $\theta_{\rm r}$ values cluster around either 0\degree\ or 90\degree\ as expected, except at the short and long ends of the wavelength considered where grism efficiencies are lower and at the wavelengths of telluric absorption bands (e.g. O$_{\rm 2}$ absorption at $\sim$760 nm). 

\section{Results \label{sec:res}}

This section examines the wavelength dependence of polarization $P_{\rm r}(\lambda)$ for our 12 C-complex target asteroids. With 52 more asteroids having V-band polarization data from polarimetric surveys, we analyze the similarities and differences between their phase-angle dependence $P_{\rm r}(\alpha)$.

\subsection{Polarimetric dependence on wavelength \label{sec:res1}}

Figure \ref{Fig02} presents reduced polarization $P_{\rm r}$ and its position angle $\theta_{\rm r}$ (Eqs. \ref{eq:eq1} and \ref{eq:eq2}) of our target asteroids as a function of wavelength. Each panel has a different y-axis scale. We rebinned data at $\sim$3 nm intervals. The lowest-resolution data ($\Delta\lambda$ $\sim$ 20--40 nm) is highlighted with salmon-colored diamond symbols. The sharp deviations of $P_{\rm r}$ and $\theta_{\rm r}$ around telluric absorption regions in the finely-binned data are absent in the coarsest data. $P_{\rm r}$ exhibits neither abrupt absorption nor emission features except near both ends of the wavelength regions and rather varies smoothly. The overall slope of $P_{\rm r}(\lambda)$ in the NPB is either neutral to negative, which is different even for some asteroids in the same taxonomic type that have similar VNIR reflectance spectra. There is no apparent correlation between the slope and the phase angle of observations. 
Interestingly, two epochs (C-type (35) Leukothea at $\alpha$ = 23\fdg4 and Ch-type (404) Arsinoe at $\alpha$ = 23\fdg4) in the PPB have a moderately positive slope and one of T-type (308) Polyxo at $\alpha$ = 20\fdg8 appears to be at the very beginning of a slope reversal, in contrast to their trends in the NPB.
A similar transition -- from a negative-to-neutral gradient in the NPB to a slightly positive gradient in the PPB -- has also been observed for intermediate-albedo asteroids \citep{Bagnulo2015}. 
We will focus on the observations in the NPB throughout the study to characterize the scattering environment of asteroids dominated by coherent backscattering; however, observations over a broader $\alpha$ (covering both NPB and PPB) should reveal differences and similarities among the asteroids in greater detail, which will be the topic of future work.

To compare the shape of polarization spectra, we took absolute values of each spectrum $|P_{\rm r}(\lambda)|$ and normalized the curve at 550 nm. However, the direct comparison of asteroids observed at different $\alpha$ should be done with caution. An asteroid's polarization spectrum would not reverse its slope (from positive to negative or vice versa) and likely be maintained in the NPB, such as (1) Ceres (panel a in Fig. \ref{Fig02}) and some asteroids discussed in \citet{Bagnulo2015}, though a slight slope variation may occur as $\alpha$ changes.
For instance, one epoch of (308) Polyxo at $\alpha$ = 20\fdg8 near the inversion angle exhibits a neutral slope (panel i in Fig. \ref{Fig02}) compared to the trend observed at $\alpha$ = 15\fdg8. 
The slope's turn to naught as the NPB nears the inversion angle may be a natural consequence of the general shape of the asteroids' NPB: the NPB has a bowl shape with a minimum polarization degree $P_{\rm min}$ (i.e., maximum $|P_{\rm r}|$) at $\alpha_{\rm min}$ $\sim$ 10\degree\ and smaller $|P_{\rm r}|$ values at other phases with zero at the inversion angle \citep{Cellino2015a}. 
As such, lumping together asteroids of different $\alpha$ facilitates qualitative analysis of trends in polarization spectrum slopes, but their quantitative differences would be only temporary. 
This is in agreement with the assumption of \citet{Bagnulo2015} that the $\alpha$ difference between the asteroids may not significantly affect, at least to the first order, the interpretation of the normalized polarimetric spectra in the NPB.

\begin{figure}[!b]
\centering
\includegraphics[width=8.2cm]{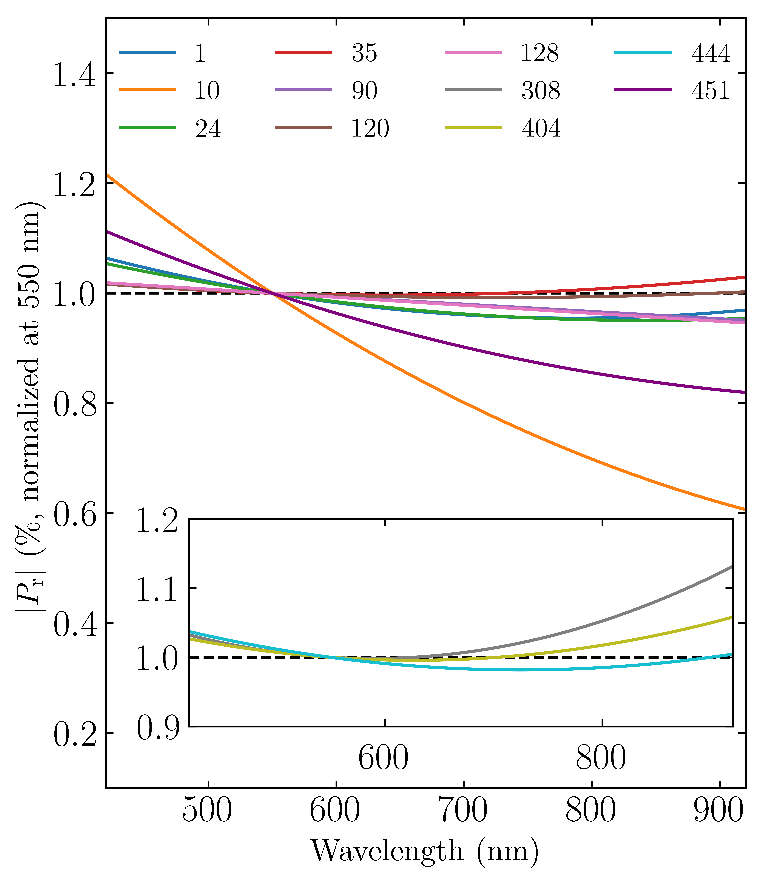}
\caption{Best-fitting quadratic curves of polarization spectra of the target asteroids normalized at 550 nm. Three asteroids showing concave shapes -- (308) Polyxo, (404) Arsinoe, and (444) Gyptis -- are visualized in a separate window. }
\label{Fig03}
\end{figure}

We excluded the data taken around the inversion angle because of their neutral slope with near-zero $P_{\rm r}$ (e.g. (308) Polyxo at $\alpha$ = 20\fdg8 and (444) Gyptis at $\alpha$ = 21\fdg2) and data of (175) Andromache due to its wavy structure that is hardly interpretable in the normalized spectrum. As a result, we sorted out 11 normalized spectra (one epoch for each asteroid) for comparison. A quadratic function was used to fit the moderately binned ($\Delta\lambda$ $\sim$ 3--5.2 nm) $|P_{\rm r}|$ spectra over 450--900 nm. For asteroids showing a strong feature around both ends of the wavelength (probably an artifact associated with low signals), particularly (128) Nemesis whose errors exceed 50 \% at wavelengths beyond 750 nm (panel g in Fig. \ref{Fig02}), we only examined data between 450 and 750 nm. 
Figure \ref{Fig03} illustrates the modeled curves for each asteroid over 420--930 nm based on their best-fitting solutions.

We here define a `polarization color' as the slope of a polarization spectrum and label it as blue (red) when the absolute $P_{\rm r}$ decreases (increases) as wavelength increases. Accordingly, an asteroid showing a positive (negative) slope in Figure \ref{Fig02} has a blue (red) polarization color. 
In Figure \ref{Fig03}, the asteroids display a (quasi-)linear slope whose color ranges from gray (0.01$\pm$0.01 \% per 100 nm) to blue ($-$0.12$\pm$0.01 \% per 100 nm) over 500--800 nm, though most vary within $\pm$5 \% of the normalized point at 550 nm. (10) Hygiea and (451) Patientia stand out for their blue polarization colors varying larger than 20 \% over the entire wavelength considered, whilst those of four outer-belt asteroids -- (24) Themis, (90) Antiope, (120) Lachesis, and (128) Nemesis -- is hardly distinguishable. 

Three asteroids (given separately in an inserted window in Fig. \ref{Fig03}) have somewhat concave shapes with a local $|P_{\rm r}|$ minimum between 600 and 720 nm: T-type (308) Polyxo, Ch-type (404) Arsinoe, and C-type (444) Gyptis in descending order of concavity. 
The concavity is always present, regardless of the spectral resolution of the data used for fitting, and leads to a change in slope, creating stronger polarization in the red part of the wavelength. The wavelength position of the local $|P_{\rm r}|$ minimum for (308) Polyxo appears consistently at a shorter wavelength around 600 nm when compared to those of the other two asteroids around 700--720 nm. 
It is interesting to note that this spectral region (the so-called 0.7-$\mu$m region; \citealt{Rivkin2002}) has been widely employed as a diagnostic for a Fe$^{\rm 2+}$--Fe$^{\rm 3+}$ charge-transfer band of phyllosilicates, a product of aqueous alteration \citep{Vilas1989,Vilas1994}.
The three asteroids with less polarized signals in this region have indeed in common that their reflectance spectra contain a measurable 0.7-$\mu$m band but also a sharp band at $\sim$2.7--2.9 $\mu$m, making them belong to the phyllosilicate-dominant `sharp' group \citep{Takir2012}. 
Furthermore, the fact that large T-type asteroids, including (308) Polyxo, possess a 0.7-$\mu$m band centered around 600--650 nm \citep{Kwon2022b}, a shorter side than that of ordinary aqueous-altered asteroids centered around 700--750 nm \citep{Fornasier2014,Rivkin2015b}, seems well compatible with the observed difference in the local $|P_{\rm r}|$ minimum wavelength of the three asteroids in Figure \ref{Fig03}.
The asteroids displaying linear polarization spectra are devoid of a corresponding infrared absorption feature and belong to the `non-sharp' group (\citealt{Rivkin2022} and references therein).

Systematic differences in the porosity (or rugosity) of the regoliths of the asteroids could affect the observed diversity in the linear slope of their polarization spectra. As dust porosity determines the effective size of working units, regoliths of different porosities will exhibit a difference in light scattering cross-sections \citep{Kolokolova2011}. In the case of large low-albedo asteroids, fine-grained ($\sim$1--10 $\mu$m) regoliths tend to exhibit more red (positive) spectral slopes in reflectance spectra than larger ($\gtrsim$10--100 $\mu$m) regoliths (e.g. \citealt{Emery2022,Emery2004}). The observed variation in slope may correspond to variations in the regolith environment that are distributed across a range of sizes if the amount of scattered light from the surface (albedo) is linearly related to the degree of polarization at a given wavelength on our sample of asteroids.

We find no correlation between the shape of polarization spectra (in terms of negativity and curvature) and phase angles, diameters, albedos, and semimajor axes of the asteroids, whose most widely accepted values are summarized in Table \ref{Fig03}. The optical spectropolarimetric properties of the large C-complex asteroids are therefore likely to be primarily determined by their surface composition, which is in line with recent spectropolarimetric observations of Didymos--Dimorphos binary systems conducted throughout the DART impact \citep{Bagnulo2023}.

\subsection{Polarimetric dependence on phase angle \label{sec:res2}}

\begin{table*}[!t]
\centering
\caption{Average linear polarization degrees of broadband filters on target asteroids}
\begin{tabular}{l|c|cccccccc}
\toprule
\hline
\multirow{2}{*}{Ast. Name} & $\alpha$ & $P_{\rm r, B}$ & $\sigma_{P_{\rm r, B}}$ & $P_{\rm r, V}$ & $\sigma_{P_{\rm r, V}}$ & $P_{\rm r, R}$ & $\sigma_{P_{\rm r, R}}$ & $P_{\rm r, I}$ & $\sigma_{P_{\rm r, I}}$ \\
 & ($^{\rm \circ}$) & (\%) & (\%) & (\%) & (\%) & (\%) & (\%) & (\%) & (\%) \\
\hline
\hline
\multirow{2}{*}{(1) Ceres} & 12.4 & $-$1.24 & 0.04 & $-$1.25 & 0.06 & $-$1.15 & 0.14 & $-$1.14 & 0.09 \\
 & 14.5 & $-$0.90 & 0.12 & $-$0.87 & 0.11 & $-$0.82 & 0.10 & $-$0.82 & 0.14 \\
\cline{2-10}
(10) Hygiea & 15.9 & $-$0.64 & 0.17 & $-$0.55 & 0.16 & $-$0.49 & 0.14 & $-$0.38 & 0.15 \\
\cline{2-10}
(24) Themis & 14.0 & $-$1.29 & 0.13 & $-$1.27 & 0.10 & $-$1.23 & 0.09 & $-$1.19 & 0.12 \\
\cline{2-10}
\multirow{2}{*}{(35) Leukothea} & 8.9 & $-$1.50 & 0.14 & $-$1.56 & 0.13 & $-$1.50 & 0.12 & $-$1.53 & 0.14 \\
 & 23.4 & 0.81 & 0.20 & 0.81 & 0.15 & 0.93 & 0.13 & 0.99 & 0.17 \\
\cline{2-10}
(90) Antiope & 8.8 & $-$1.48 & 0.18 & $-$1.46 & 0.14 & $-$1.41 & 0.15 & $-$1.40 & 0.19\\
\cline{2-10}
(120) Lachesis & 10.9 & $-$1.45 & 0.19 & $-$1.41 & 0.16 & $-$1.42 & 0.15 & $-$1.41 & 0.17 \\
\cline{2-10}
(128) Nemesis & 9.0 & $-$1.53 & 0.21 & $-$1.50 & 0.16 & $-$1.47 & 0.16 & $-$1.49 & 0.19 \\
\cline{2-10}
(175) Andromache & 16.4 & $-$0.41 & 0.19 & $-$0.36 & 0.09 & $-$0.37 & 0.09 & $-$0.36 & 0.12 \\
\cline{2-10}
\multirow{2}{*}{(308) Polyxo} & 15.8 & $-$1.17 & 0.19 & $-$1.13 & 0.17 & $-$1.13 & 0.15 & $-$1.23 & 0.21 \\
 & 20.8 & $-$0.10 & 0.11 & $-$0.01 & 0.10 & 0.00 & 0.07 & 0.00 & 0.15 \\
\cline{2-10}
\multirow{2}{*}{(404) Arsinoe} & 7.6 & $-$1.33 & 0.19 & $-$1.35 & 0.17 & $-$1.31 & 0.18 & $-$1.35 & 0.18 \\
 & 23.4 & 0.81 & 0.17 & 0.86 & 0.15 & 0.97 & 0.14 & 1.04 & 0.16 \\
\cline{2-10}
\multirow{2}{*}{(444) Gyptis} & 7.0 & $-$1.10 & 0.15 & $-$1.07 & 0.15 & $-$1.07 & 0.12 & $-$1.06 & 0.18 \\
 & 21.2 & $-$0.31 & 0.26 & $-$0.23 & 0.12 & $-$0.17 & 0.28 & $-$0.19 & 0.20 \\
\cline{2-10}
(451) Patientia & 12.4 & $-$1.54 & 0.20 & $-$1.45 & 0.15 & $-$1.32 & 0.16 & $-$1.22 & 0.22 \\
\hline
\bottomrule
\end{tabular}
\tablefoot{\centering Degree of linear polarization $P_{\rm r, X}$ and its error $\sigma_{P_{\rm r, X}}$ were measured by weight-averaging data points in the filter domain X. 
} 
\label{t2}
\end{table*}

Asteroids draw a unique bell-shaped polarimetric-phase curve $P_{\rm r}(\alpha)$  that can be described as follows: as a phase angle $\alpha$ increases from zero, $P_{\rm r}$ becomes more negative, reaches its minimum $P_{\rm min}$ at $\alpha_{\rm min}$ of $\sim$5--15\degree, and then rebounds to zero $P_{\rm r}$ at the inversion angle $\alpha_{\rm 0}$ of $\sim$15--30\degree\ beyond which $P_{\rm r}$ becomes positive \citep{Cellino2015a,Belskaya2017}. Two chief parameters defining the shape of the NPB -- $\alpha_{\rm 0}$ (the angular width of the NPB) and $P_{\rm min}$ (the depth of the NPB, usually in absolute number) -- are of particular importance thanks to their sensitivity to the regolith's compositional and physical properties \citep{Dollfus1975,Geake1986,Belskaya2005}.

\begin{figure}[!t]
\centering
\includegraphics[width=8.2cm]{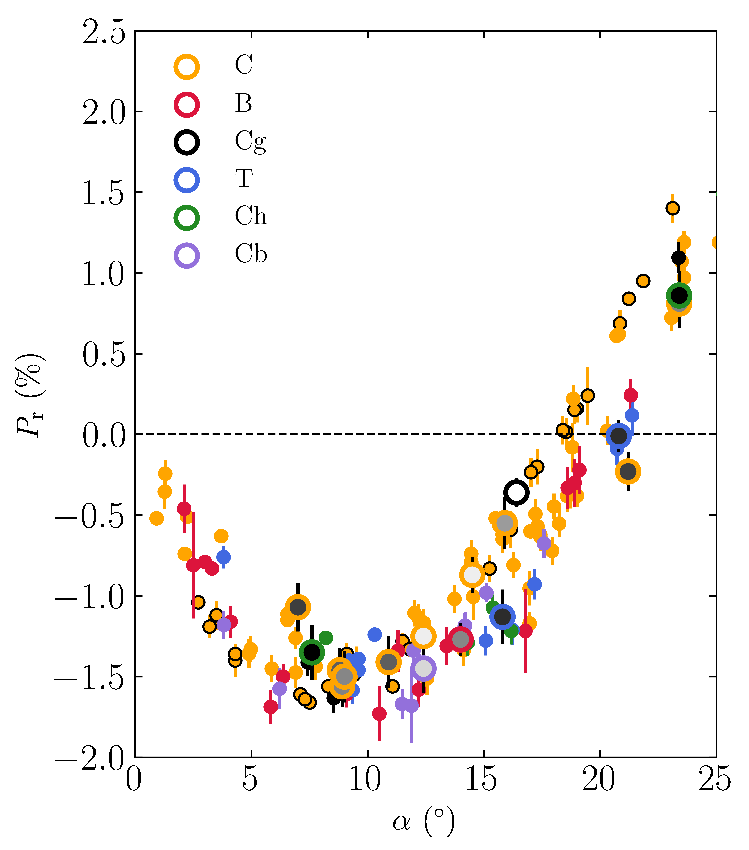}
\caption{V-band polarization-phase curve $P_{\rm r}(\alpha)$ of our 12 target asteroids. Symbols with thick line edges represent the data measured from this study, with gray-scale colors indicating their relative albedos ranging from white (high-albedo end: 0.094 of (175) Andromache) and black (low-albedo end: 0.037 of (404) Arsinoe). The albedos were from the NASA JPL Small-Body Database (\protect\url{https://ssd.jpl.nasa.gov/tools/sbdb\_lookup.html\#/}).
Background circles are their archival data from the NASA Planetary Data System (PDS) Small Bodies Node \citep{Lupishko2022} and Calern Asteroid Polarization Survey (CAPS; \citealt{Bendjoya2022}). Points of the dwarf planet (1) Ceres are distinguished by black-edged orange circles.}
\label{Fig04}
\end{figure}

To grasp the $P_{\rm r}(\alpha)$ of our C-complex asteroids in the first place, we averaged polarization data bracketed in the standard B (405--485 nm), V (510--590 nm), R (620--670 nm), and I (730--880 nm) band regions. Table \ref{t2} summarizes the weighted-average $P_{\rm r, X}$ and its error $\sigma_{P_{\rm r, X}}$ in each wavelength domain X. As for the B and I bands, the presence of deviating structures in polarization spectra as signals weaken (Fig. \ref{Fig02}) makes the data less reliable than the values of the V and R bands. As archival V-band polarimetric data significantly outnumber those of other bands, we utilized V-band data for the analysis. Figure \ref{Fig04} shows the polarization-phase curves of our targets in the V band. Data points from this study (Table \ref{t2}) are highlighted as gray-scale symbols (white and black indicate the brightest and darkest ends of (175) Andromache and (404) Arsinoe, respectively), while the background circles represent their archival data quoted from the NASA Planetary Data System (PDS) Small Bodies Node \citep{Lupishko2022} and Calern Asteroid Polarization Survey (CAPS; \citealt{Bendjoya2022}). 

The $P_{\rm r, V}$ values from our spectropolarimetric results share a consistent phase-angle dependence with the existing datasets derived from aperture polarimetry. Despite the different subclasses, the C-complex asteroids draw a broadly homogeneous $P_{\rm r}(\alpha)$, as shown in \citet{Belskaya2017}. The data points show no obvious correlation between the asteroids' albedo and relative $P_{\rm r}$ excess at the given $\alpha$, although, in general, the more diffuse light component contributes to the observed scattered light (i.e., higher-albedo surfaces), the easier it is to lose the directional information and become depolarized. A lack of such an inverse relationship between albedo and polarization (so-called the Umov law; \citealt{Umov1905}) in our C-complex asteroids, or at least in the NPB we are exploring, is consistent with previous laboratory studies showing that the Umov law becomes saturated at an albedo of $\sim$0.08 \citep{Zellner1977a,Zellner1977b,Geake1986,Cellino2015b}.

We next increased the number of large C-complex asteroids from NASA/PDS and CAPS archives in order to parameterize the NPB phase curve using a conventional trigonometric function \citep{Lumme1993}:
\begin{equation}
P_{\rm r}(\alpha) = b~\sin^{c_{\rm 1}}(\alpha)~\cos^{c_{\rm 2}}\Bigg(\frac{\alpha}{2}\Bigg)~\sin(\alpha - \alpha_{\rm 0})
\label{eq:eq3}
\end{equation}
\noindent or through the linear-exponential function \citep{Kaasalainen2003}:
\begin{equation}
P_{\rm r}(\alpha) = A~({\rm e}^{-(\alpha/B)} - 1) + C\alpha~,
\label{eq:eq4}
\end{equation}
\noindent where constants $b$, $c_{\rm 1}$, $c_{\rm 1}$, and $\alpha_{\rm 0}$ (inversion angle) in Eq. \ref{eq:eq3} and $A$, $B$, and $C$ in Eq. \ref{eq:eq4} are free parameters shaping the curve, retrieved by best-fit techniques. With these two fitting functions in hand, we only considered asteroids if they simultaneously satisfy the criteria listed below:
\begin{enumerate}
\item[(1)] large ($D_{\rm e}$ $>$ 100 km) C-complex asteroids;
\item[(2)] a minimum of four V-band measurements including the point (0,0), that is, $P_{\rm r}$ = 0 \% at $\alpha$ = 0$^{\circ}$, assuming that there is no significant alignment of regolith particles;
\item[(3)] $\alpha$ distributed evenly covering more than 10\degree\ in the NPB;
\item[(4)] agreement of the $\alpha_{\rm 0}$ and $P_{\rm min}$ values retrieved from the trigonometric and linear-exponential fitting functions within the error bars;
\item[(5)] a fitting uncertainty of $<$20 \%.  
\end{enumerate}

Condition (4) arises from the empirical nature of both fitting functions which makes judging one over the other pointless. This constraint is usually met if there are sufficient observations available. When compared to Eq. \ref{eq:eq3}, which is capable of fitting polarization-phase curves up to high phases, Eq. \ref{eq:eq4} generally fits the curves of main-belt asteroids whose $\alpha$-coverage hardly exceeds 30\degree.
The final estimates were selected based upon the conventional trigonometric function (Eq. \ref{eq:eq3}), though we are dealing with main-belt asteroids. This is because within the limit of 5 000 iterations we arbitrarily set, Eq. \ref{eq:eq3} tends to provide more stable error estimates than Eq. \ref{eq:eq4}, possibly due to a lack of dense coverage at small $\alpha$ as pointed by \citet{Kaasalainen2003}. Consequently, 64 asteroids were selected, whose retrieved polarimetric parameters with 1$\sigma$ errors are available in Table \ref{t3} along with their taxonomic types. In some cases, C-complex asteroids have more than one subclass or are classified differently under Tholen \citep{Tholen1984} and Bus-DeMeo systems (SMASSII; \citealt{Bus2002,DeMeo2009}). In Table \ref{t3}, the selected taxonomic type is highlighted in boldface and the reasons for selection are given in Appendix \ref{sec:app1}.

\begin{figure}[!b]
\centering
\includegraphics[width=8.5cm]{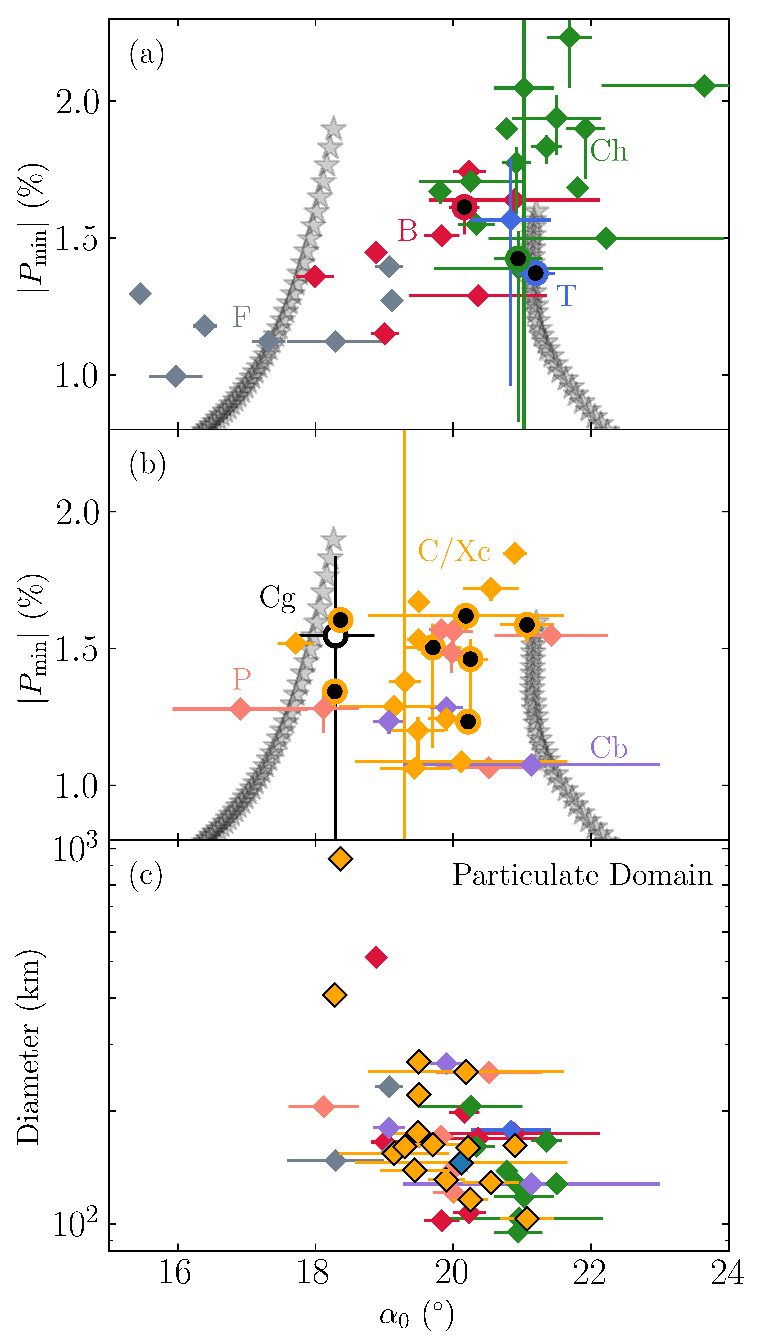}
\caption{Distribution of 64 large C-complex asteroids in the retrieved inversion angle $\alpha_{\rm 0}$ and absolute polarization minimum $|P_{\rm min}|$ space (panels a and b) and the size of the asteroids (panel c). Guidelines in panels a and b represent the borders of surface regimes dominated by different-size regolith particles \citep{Geake1986}: the left border separates the boulder/bare-rock regoliths (leftmost) and particular surface domain (intermediate) and the right one divides the particulate domain and fine-dust regoliths (rightmost). Different colors are assigned to different taxonomic types.
Panel c illustrates the distribution of asteroids in the particulate domain.}
\label{Fig05}
\end{figure}

Figure \ref{Fig05} summarizes the fitting results. A pair of border lines measured from lunar samples \citep{Geake1986,Dollfus1989} separate regolith regimes into three domains based on $\alpha_{\rm 0}$ values: bare-rock/bouldery (leftmost), particulate (intermediate), and fine-dust/fluffy (rightmost) regoliths.
The $|P_{\rm min}|$ range of the asteroids is consistent with that of carbonaceous meteorites \citep{Zellner1977a,Zellner1977b}. Figures \ref{Fig05}a and \ref{Fig05}b displays the distribution of asteroids having different taxonomic types in the retrieved $\alpha_{\rm 0}$--$|P_{\rm min}|$ space. The distribution of asteroids (F, B, T, and Ch types) in panel a illustrates the localization of each subgroup within the parameter space, while in panel b, taxonomic types are either distributed throughout the entire parameter range (P and C/Xc types) or are lacking sufficient data to draw any meaningful conclusions (Cg and Cb types).

The $\alpha_{\rm 0}$ and $|P_{\rm min}|$ of F-, B-, T-, and Ch-type asteroids in Figure \ref{Fig05}a are proportional, and the stream aligns with the direction of positive correlation (i.e., the stronger the negative polarization, the wider the NPB) in ascending order of F (weighted mean and its 1-$\sigma$ error: $\alpha_{\rm 0}$ = 17\fdg10$\pm$1\fdg64 and $|P_{\rm min}|$ = 1.22$\pm$0.09 \%), B ($\alpha_{\rm 0}$ = 19\fdg04$\pm$0\fdg48 and $|P_{\rm min}|$ = 1.40$\pm$0.16 \%), T ($\alpha_{\rm 0}$ = 21\fdg14$\pm$0\fdg14 and $|P_{\rm min}|$ = 1.37$\pm$0.01 \%), and Ch ($\alpha_{\rm 0}$ = 21\fdg07$\pm$0\fdg47 and $|P_{\rm min}|$ = 1.76$\pm$0.19 \%) types. The Spearman correlation coefficient $\rho$ and its p-value of the distribution as a whole are $\rho$ = 0.70$^{+0.06}_{-0.07}$ and $p$ = 0.0001, respectively. The statistical details are provided in Appendix \ref{sec:app2}.
F-type asteroids occupy the smaller end of $\alpha_{\rm 0}$ compared to most of the C-complex asteroids, as noted in previous studies \citep{Belskaya2005,Belskaya2017}. 

It is intriguing to see that the sequence of distribution (F $\rightarrow$ B $\rightarrow$ T $\rightarrow$ Ch) in the parameter space appears to parallel that of the expected degree of aqueous alteration (P $<$ B $<$ T $\lesssim$ Ch, no F types in the samples) based on the detection rate of a 0.7-$\mu$m band and the depth of a 2.9-$\mu$m band depth \citep{Fornasier2014,Kwon2022b}. About half of the F- and Ch-type points reside outside the particulate domain where most large asteroids are expected to position \citep{Marchis2012}. We will return to these topics in Section \ref{sec:discuss}.

C/Xc-type asteroids distribute across the entire particulate domain (Fig. \ref{Fig05}b). This apparent diversity indicates that this taxonomic group may indeed have highly heterogeneous surfaces and/or may belong to another C-complex subgroup but possess too weak diagnostic features to be detected. 

\onecolumn
\begin{small}
\centering
\begin{ThreePartTable}
\begin{TableNotes}
\footnotesize
\item [a] Visible geometric albedo
\item [b] Effective diameters in kilometers. $p_{\rm V}$ and $D_{\rm e}$ are quoted from JPL Small-Body Database (https://ssd.jpl.nasa.gov/tools/sbdb\_lookup.html\#/).
\item [c] Semimajor axis. The main asteroid belt is divided into three zones based on the semimajor axis: inner ($a$ $<$ 2.502 au), middle (2.502 au $\le$ $a$ $\le$ 2.825 au), and outer ($a$ $>$ 2.825 au)
\item [d] Tholen \citep{Tholen1984} and Bus-DeMeo \citep{DeMeo2009} taxonomies. When the two systems classify an object into different classes, C types have the lowest priority. The selected taxonomic type is highlighted in bold. Asteroids footnoted with stars (${\star}$) were re-classified in this study for consistent analysis based on our selection criteria (Appendix \ref{sec:app1}).
\item [e] Inversion angle in degrees
\item [f] Polarization minimum in absolute numbers in percent
\item [g] Modified bend parameter (Eq. \ref{eq:eq5}). The original expression was suggested by \citet{McCord1975a,McCord1975b}.
\item [h] Band center in micrometers. As in the reference studies, we use micrometer units for the 0.7-$\mu$m and 3-$\mu$m band properties. `$N$' and `$-$' denote non-detection and no available data, respectively.
\item [i] 3-$\mu$m band classification criteria following \citet{Rivkin2022}. `U' denotes unclear. \\
{\bf References.} (1) \citet{Fornasier1999}; (2) \citet{Rivkin2019}; (3) \citet{Rivkin2015b}; (4) \citet{Fornasier2014}; (5) \citet{Takir2012}; (6) \citet{Rivkin2022}; (7) \citet{Usui2019}; (8) This study; (9) \citet{Yang2009}; (10) \citet{Hargrove2012}; (11) \citet{Hargrove2015}; (12) \citet{Kwon2022b}; (13) \citet{Takir2023}; (14) \citet{Hiroi1993}.
\end{TableNotes}
\begin{longtable}{c|cc|cc|cc|c|c|c|c|c|c}

\caption{Basic profiles, polarimetric and spectroscopic parameters of the asteroids used in this study and their references} \label{t3} \\

\toprule
\hline
\multirow{2}{*}{Ast.} & \multirow{2}{*}{$p_{\rm V}$\tnote{a}} & $D_{\rm e}$\tnote{b} & \multirow{2}{*}{$a$\tnote{c}} & \multirow{2}{*}{Taxon.\tnote{d}} & $\alpha_{\rm 0}$\tnote{e} & |$P_{\rm min}$|\tnote{f} & \multirow{2}{*}{Bend\tnote{g}} & \multicolumn{2}{c}{0.7-$\mu$m band} & \multicolumn{1}{|c|}{3-$\mu$m} & \multicolumn{2}{c}{Reference} \\
\cline{9-10}
\cline{12-13}
 & & (km) & & & ($^{\rm \circ}$) & (\%) & & $\lambda_{\rm c}$\tnote{h} ($\mu$m) & Depth & Group\tnote{i} & 0.7 $\mu$m & 3 $\mu$m \\ 
\hline
\hline
\endfirsthead

\caption[]{Continued}\\
\toprule
\hline
\multirow{2}{*}{Ast.} & \multirow{2}{*}{$p_{\rm V}$$^{\rm a}$} & $D_{\rm e}$$^{\rm b}$ & \multirow{2}{*}{$a$$^{\rm c}$} & \multirow{2}{*}{Taxon.$^{\rm d}$} & $\alpha_{\rm 0}$$^{\rm e}$ & |$P_{\rm min}$|$^{\rm f}$ & \multirow{2}{*}{Bend$^{\rm g}$} & \multicolumn{2}{c}{0.7-$\mu$m band} & \multicolumn{1}{|c|}{3-$\mu$m} & \multicolumn{2}{c}{Reference} \\
\cline{9-10}
\cline{12-13}
 & & (km) & & & ($^{\rm \circ}$) & (\%) & & $\lambda_{\rm c}$$^{\rm h}$ ($\mu$m) & Depth & Group$^{\rm i}$ & 0.7 $\mu$m & 3 $\mu$m \\ 
\hline
\hline
\endhead


\bottomrule
\endfoot

\bottomrule
\insertTableNotes  
\endlastfoot

 1 & 0.090 & 939.4 & middle & G/{\bf C} & 18.36$\pm$0.09 & 1.60$^{+0.02}_{-0.03}$ & 0.05$\pm$0.01 & N & N & NST & 1 & 2 \\
 2 & 0.155 & 513 & middle & B/{\bf B} & 18.88$\pm$0.06 & 1.45$^{+0.01}_{-0.01}$ & 0.01$\pm$0.04 & $-$ & $-$ & ST & $-$ & 3 \\
 10 & 0.072 & 407.12 & outer & C/{\bf C} & 18.28$\pm$0.09 & 1.34$^{+0.05}_{-0.07}$ & 0.06$\pm$0.01 & $\sim$0.66 & 1.0$\pm$0.1 & NST & 4 & 2 \\
 13 & 0.049 & 202.636 & middle & G/{\bf Ch} & 21.69$\pm$0.33 & 2.23$^{+0.07}_{-0.19}$ & 0.11$\pm$0.03 & $\sim$0.68 & 2.4$\pm$0.1 & ST & 4 & 3 \\
 24 & 0.067 & 198 & outer & C/{\bf B} & 20.16$\pm$0.22 & 1.61$^{+0.04}_{-0.10}$ & 0.05$\pm$0.01 & $\sim$0.67 & 3.5$\pm$0.1 & NST & 1 & 5 \\
 31 & 0.053 & 267.080 & outer & C/{\bf Cb} & 19.90$\pm$0.24 & 1.28$^{+0.01}_{-0.02}$ & 0.04$\pm$0.02 & N & N & NST & 4 & 2 \\
 35 & 0.066 & 103.055 & outer & C/{\bf C} & 21.07$\pm$0.39 & 1.59$^{+0.04}_{-0.06}$ & 0.12$\pm$0.01 &  $-$ & $-$ &  $-$ &  $-$ &  $-$\\
 41 & 0.059 & 205.495 & middle & C/{\bf Ch} & 20.25$\pm$0.75 & 1.71$^{+0.02}_{-0.02}$ & 0.11$\pm$0.02 & $\sim$0.69 & 3.3$\pm$0.2 & ST & 1 & 3 \\
 45 & 0.045 & 202.327 & middle & {\bf F}C/C & 16.40$\pm$0.17 & 1.18$^{+0.02}_{-0.02}$ & 0.03$\pm$0.02 & N & N & NST & 1 & 6 \\
 46 & 0.046 & 131.471 & middle & {\bf Xc$^{\star}$} & 17.71$\pm$0.27 & 1.52$^{+0.02}_{-0.04}$ & $-$0.01$\pm$0.01 & $-$ & $-$ & U & $-$ & 7 \\
 47 & 0.082 & 168.174 & outer & C/{\bf B} & 17.99$\pm$0.27 & 1.36$^{+0.02}_{-0.02}$ & 0.03$\pm$0.03 & N & N & $-$ & 4 & $-$ \\
 48 & 0.065 & 216.473 & outer & CG/{\bf Ch} & 23.65$\pm$1.50 & 2.06$^{+0.02}_{-0.02}$ & 0.12$\pm$0.01 & $\sim$0.68 & 2.3$\pm$0.1 & ST & 4 & 3 \\
 49 & 0.048 & 166.252 & outer & CG/{\bf Ch} & 21.35$\pm$0.22 & 1.83$^{+0.04}_{-0.06}$ & 0.15$\pm$0.03 & $\sim$0.71 & 5.4$\pm$0.4 & ST & 8 & 7 \\
 51 & 0.100 & 138.159 & inner & CU/{\bf Ch} & 20.78$\pm$0.07 & 1.90$^{+0.01}_{-0.01}$ & 0.09$\pm$0.01 & $\sim$0.71 & 5.3$\pm$0.2 & ST & 1 & 3 \\
 54 & 0.059 & 160.120 & middle & {\bf Ch$^{\star}$} & 21.92$\pm$0.28 & 1.90$^{+0.01}_{-0.18}$ & 0.11$\pm$0.03 & $\sim$0.68 & 3.9$\pm$0.1 & ST & 4 & 5 \\
 56 & 0.057 & 121.333 & middle & {\bf P$^{\star}$} & 20.00$\pm$0.29 & 1.56$^{+0.01}_{-0.07}$ & 0.04$\pm$0.02 & N & N & ST & 4 & 7 \\
 58 & 0.044 & 106.517 & middle & C/{\bf Ch} & 22.22$\pm$1.71 & 1.50$^{+0.01}_{-0.02}$ & 0.08$\pm$0.01 & $\sim$0.71 & 2.4$\pm$0.1 & $-$ & 4 & $-$ \\
 59 & 0.044 & 165.119 & middle & CP/{\bf B} & 19.01$\pm$0.20 & 1.15$^{+0.01}_{-0.02}$ & $-$0.05$\pm$0.01 & $-$ & $-$ & NST & $-$ & 6 \\
 62 & 0.048 & 106.921 & outer & {\bf B}U/Ch & 20.23$\pm$0.24 & 1.74$^{+0.02}_{-0.07}$ & 0.05$\pm$0.03 & N & N & $-$ & 9 & $-$ \\
 76 & 0.058 & 145.423 & outer & {\bf Xc$^{\star}$} & 20.12$\pm$1.5 & 1.09$^{+0.01}_{-0.03}$ & 0.04$\pm$0.03 & $-$ & $-$ & NST & $-$ & 10 \\
 78 & 0.071 & 120.60 & middle & C/{\bf Ch} & 21.81$\pm$0.12 & 1.68$^{+0.02}_{-0.02}$ & 0.09$\pm$0.01 & $\sim$0.69 & 2.4$\pm$0.1 & ST & 4 & 3 \\
 85 & 0.067 & 154.79 & middle & {\bf F}C/B & 19.11$\pm$0.12 & 1.27$^{+0.01}_{-0.01}$ & 0.02$\pm$0.01 & N & N & $-$ & 4 & $-$ \\
 87 & 0.046 & 253.051 & outer & {\bf P$^{\star}$} & 20.52$\pm$0.77 & 1.06$^{+0.01}_{-0.01}$ & 0.03$\pm$0.04 & $-$ & $-$ & U & $-$ & 7 \\
 88 & 0.067 & 232 & middle & C{\bf F}/B & 19.07$\pm$0.20 & 1.39$^{+0.02}_{-0.03}$ & 0.02$\pm$0.01 & $-$ & $-$ & NST & $-$ & 2 \\
 90 & 0.058 & 115.974 & outer & C/{\bf C} & 20.25$\pm$0.26 & 1.46$^{+0.07}_{-0.25}$ & 0.04$\pm$0.02 & $\sim$0.69 & 0.8$\pm$0.1 & NST & 4 & 11 \\
 91 & 0.048 & 103.402 & middle & CP/{\bf Ch} & 20.95$\pm$1.2 & 1.39$^{+0.14}_{-0.56}$ & 0.11$\pm$0.01 & $\sim$0.71 & 2.9$\pm$0.4 & ST & 8 & 5 \\
 93 & 0.056 & 154.155 & middle & CU/{\bf C} & 19.14$\pm$0.80 & 1.29$^{+0.01}_{-0.04}$ & 0.05$\pm$0.02 & N & N & NST & 4 & 6 \\
 94 & 0.040 & 204.89 & outer & C{\bf P}/C & 18.12$\pm$0.51 & 1.28$^{+0.07}_{-0.09}$ & 0.02$\pm$0.01 & $-$ & $-$ & NST & $-$ & 6 \\
 96 & 0.048 & 177.774 & outer & T/{\bf T} & 20.84$\pm$0.58 & 1.57$^{+0.23}_{-0.61}$ & $-$0.02$\pm$0.01 & $\sim$0.60 & 1.6$\pm$0.1 & ST & 12 & 12 \\
 117 & 0.027 & 208.974 & outer & XC/{\bf C} & 22.19$\pm$2.63 & 1.19$^{+0.02}_{-0.06}$ & 0.05$\pm$0.01 & $-$ & $-$ & ST & $-$ & 6 \\
 120 & 0.058 & 155.132 & outer & C/{\bf C} & $-$ & $-$ & 0.06$\pm$0.01 & N & N & ST & 4 & 5 \\ 
 128 & 0.067 & 162.515 & outer & C/{\bf C} & 19.70$\pm$0.42 & 1.50$^{+0.09}_{-0.37}$ & 0.02$\pm$0.02 & N & N & ST & 1 & 7 \\ 
 140 & 0.068 & 109.79 & middle & {\bf P$^{\star}$} & 21.42$\pm$0.83 & 1.55$^{+0.02}_{-0.02}$ & 0.00$\pm$0.01 & N & N & NST & 4 & 7 \\
 141 & 0.067 & 117.916 & middle & CPF/{\bf Ch} & 21.03$\pm$0.43 & 2.05$^{+1.10}_{-3.00}$ & 0.07$\pm$0.02 & $\sim$0.71 & 4.1$\pm$0.3 & $-$ & 8 & $-$ \\
 145 & 0.061 & 127.783 & middle & C/{\bf Ch} & 21.50$\pm$0.65 & 1.94$^{+0.08}_{-0.14}$ & 0.07$\pm$0.02 & $\sim$0.70 & 3.6$\pm$0.1 & ST & 4 & 7 \\
 153 & 0.062 & 170.63 & outer & {\bf P$^{\star}$} & 16.91$\pm$0.99 & 1.28$^{+0.01}_{-0.03}$ & 0.02$\pm$0.02 & N & N & NST & 4 & 5 \\
 165 & 0.047 & 180.083 & outer & CD/{\bf Cb} & 19.07$\pm$0.23 & 1.23$^{+0.02}_{-0.05}$ & 0.03$\pm$0.01 & $-$ & $-$ & NST & $-$ & 13 \\
 175 & 0.094 & 94.532 & outer & C/{\bf Cg} & 18.29$\pm$0.56 & 1.55$^{+0.29}_{-0.87}$ & 0.17$\pm$0.03 & N & N & NST & 4 & 6 \\
 185 & 0.062 & 160.464 & middle & C/{\bf C} & 19.30$\pm$0.23 & 1.38$^{+1.01}_{-2.38}$ & 0.04$\pm$0.02 & N & N & NST & 4 & 6 \\
 194 & 0.057 & 161.667 & middle & C/{\bf C} & 20.90$\pm$0.17 & 1.85$^{+0.01}_{-0.01}$ & 0.08$\pm$0.01 & $\sim$0.69 & 2.7$\pm$0.1 & $-$ & 4 & $-$ \\
 200 & 0.053 & 128.301 & middle & C/{\bf Ch} & 20.91$\pm$0.21 & 1.77$^{+0.06}_{-0.19}$ & 0.06$\pm$0.02 & $\sim$0.70 & 2.2$\pm$0.3 & ST & 1 & 3 \\
 241 & 0.058 & 168.90 & outer &  CP/{\bf B} & 20.36$\pm$1.00 & 1.29$^{+0.01}_{-0.03}$ & $-$0.01$\pm$0.01 & $-$ & $-$ & NST & $-$ & 6 \\
 247 & 0.064 & 130.935 & middle & {\bf Xc$^{\star}$} & 19.90$\pm$0.27 & 1.24$^{+0.01}_{-0.05}$ & 0.05$\pm$0.02 & $-$ & $-$ & NST & $-$ & 6 \\
 259 & 0.043 & 174.318 & outer & {\bf Xc$^{\star}$} & 19.49$\pm$0.39 & 1.20$^{+0.05}_{-0.11}$ & 0.05$\pm$0.01 & N & N & ST & 4 & 6 \\
 308 & 0.047 & 128.578 & middle & T/{\bf T} & 21.20$\pm$0.27 & 1.37$^{+0.01}_{-0.01}$ & $-$0.01$\pm$0.01 & $\sim$0.60 & 5.0$\pm$0.2 & ST & 12 & 5 \\
 324 & 0.050 & 220.691 & middle & {\bf Xc$^{\star}$} & 19.50$\pm$0.17 & 1.53$^{+0.01}_{-0.01}$ & 0.07$\pm$0.01 & $-$ & $-$ & NST & $-$ & 2 \\
 350 & 0.048 & 128.729 & outer & {\bf C}/$-$ & 20.55$\pm$0.40 & 1.72$^{+0.01}_{-0.05}$ & 0.07$\pm$0.05 & $\sim$0.69 & 3.1$\pm$0.1 & $-$ & 4 & $-$ \\
 372 & 0.059 & 173.642 & outer & BFC/{\bf B} & 20.89$\pm$1.24 & 1.64$^{+0.01}_{-0.05}$ & 0.09$\pm$0.01 & N & N & $-$ & 4 & $-$ \\
 381 & 0.064 & 127.639 & outer & C/{\bf Cb} & 21.14$\pm$1.87 & 1.08$^{+0.03}_{-0.04}$ & 0.06$\pm$0.01 & N & N & $-$ & 4 & $-$ \\
 386 & 0.070 & 165.01 & outer & {\bf Ch$^{\star}$} & 19.81$\pm$0.11 & 1.67$^{+0.01}_{-0.05}$ & 0.08$\pm$0.01 & $\sim$0.68 & 3.3$\pm$0.1 & ST & 4 & 6 \\
 404 & 0.037 & 94.970 & middle & C/{\bf Ch} & 20.94$\pm$0.35 & 1.42$^{+0.03}_{-0.05}$ & 0.09$\pm$0.01 & $\sim$0.70 & 3.7$\pm$0.5 & ST & 8 & 3 \\
 409 & 0.054 & 171.012 & middle & {\bf P$^{\star}$} & 19.82$\pm$0.18 & 1.57$^{+0.01}_{-0.03}$ & $-$0.02$\pm$0.01 & N & N & $-$ & 4 & $-$\\
 419 & 0.034 & 148.701 & middle & {\bf F}/$-$ & 15.97$\pm$0.39 & 0.99$^{+0.02}_{-0.03}$ & 0.01$\pm$0.02 & N & N & U & 4 & 7 \\
 420 & 0.044 & 138.699 & outer & {\bf Xc$^{\star}$} & 19.44$\pm$0.51 & 1.06$^{+0.01}_{-0.03}$ & $-$ & $-$ & $-$ & $-$ & $-$ & $-$ \\
 423 & 0.067 & 175.859 & outer & C/{\bf C} & 20.87$\pm$2.19 & 1.30$^{+0.07}_{-0.21}$ & 0.07$\pm$0.01 & $-$ & $-$ & NST & $-$ & 6 \\
 431 & 0.055 & 101.900 & outer & B/{\bf B} & 19.83$\pm$0.26 & 1.51$^{+0.03}_{-0.04}$ & 0.02$\pm$0.01 & $-$ & $-$ & $-$ & $-$ & $-$ \\
 444 & 0.051 & 159.311 & middle & C/{\bf C} & 20.22$\pm$0.11 & 1.23$^{+0.01}_{-0.01}$ & 0.05$\pm$0.01 & $\sim$0.69 & 1.5$\pm$0.1 & ST & 1 & 6 \\
 451 & 0.058 & 155.132 & outer & C/{\bf C} & 20.19$\pm$1.43 & 1.62$^{+0.02}_{-0.05}$ & 0.01$\pm$0.01 & $-$ & $-$ & NST & $-$ & 2 \\ 
 476 & 0.035 & 138.493 & middle & {\bf P$^{\star}$} & 19.98$\pm$0.23 & 1.48$^{+0.03}_{-0.08}$ & 0.00$\pm$0.01 & N & N & U & 4 & 7 \\
 511 & 0.076 & 270.327 & outer & C/{\bf C} & 19.50$\pm$0.12 & 1.67$^{+0.01}_{-0.02}$ & 0.04$\pm$0.01 & $\sim$0.72 & $-$ & ST & 14 & 7 \\
 654 & 0.027 & 160.736 & inner & C/{\bf Ch} & 20.34$\pm$0.27 & 1.55$^{+0.01}_{-0.02}$ & 0.05$\pm$0.02 & $\sim$0.67 & 2.3$\pm$0.1 & ST & 4 & 3 \\
 704 & 0.078 & 306.313 & outer & {\bf F}/B & 15.46$\pm$0.10 & 1.30$^{+0.01}_{-0.01}$ & 0.04$\pm$0.01 & N & N & NST & 4 & 2 \\
 762 & 0.040 & 147.343 & outer & {\bf F}/$-$ & 18.29$\pm$0.70 & 1.12$^{+0.01}_{-0.03}$ & 0.00$\pm$0.49 & $-$ & $-$ & $-$ & $-$ & $-$ \\
 1021 & 0.045 & 100.765 & middle & {\bf F}/B & 17.32$\pm$0.25 & 1.12$^{+0.01}_{-0.02}$ & $-$0.01$\pm$0.04 & N & N & $-$ & 4 & $-$ \\
\hline
\end{longtable}
\end{ThreePartTable}
\end{small}
\twocolumn
\raggedbottom

Lastly, we find that there is a weak but notable correlation between $\alpha_{\rm 0}$ and the diameter of asteroids whose NPB parameters are within the particulate domain (Fig. \ref{Fig05}c). The asteroids analyzed here roughly align with the inverse $\alpha_{\rm 0}$--diameter correlation ($\rho$ = $-$0.42$^{+0.08}_{-0.07}$ and $p$ = 0.01; Appendix \ref{sec:app2}) as a whole and possibly within each taxonomic group, most notably in C-type asteroids  ($\rho$ = $-$0.51$^{+0.13}_{-0.11}$ and $p$ = 0.04). 
According to the $\alpha_{\rm 0}$--regolith size relationship (assuming that this bordering scheme is applicable to large C-type asteroids), the larger asteroids with smaller $\alpha_{\rm 0}$ values would have coarser regolith grains than the smaller asteroids with larger $\alpha_{\rm 0}$ values. This trend contradicts general observations, where small asteroids (order of 1--10 km) that hardly retain pulverized collisional fragments create regoliths dominated by boulders of millimeter size, whereas larger asteroids (order of 100 km) have sufficient gravity to retain smaller fragments of micrometers to tens of micrometers \citep{Dollfus1989,Bottke2005,Delbo2007,Gundlach2013}. Moreover, this $\alpha_{\rm 0}$--regolith size link seems to contradict a recent study showing that large F-type asteroids exhibit thermal emissivity properties equivalent to fine-grained regoliths as primitive Jupiter Trojans \citep{Marchis2012,Haoxuan2023}, whose smaller $\alpha_{\rm 0}$ values are, according to the bordering scheme, indicative of coarser regoliths than other C-complex asteroids.
Further observations of large asteroids, including non-C-complex asteroids, are needed to confirm this correlation.
Aside from the results shown in Figure \ref{Fig05}, we are unable to find statistically significant relationships between the polarimetric parameters and the surface characteristics (e.g. albedo) of the asteroids.

\section{Discussion \label{sec:discuss}}

This section provides polarimetric evidence for surface heterogeneity in large C-complex asteroids. Using the observed correlation of polarimetric and spectroscopic properties, we attempt to tie VNIR observations to asteroids' surface environments and discuss what future polarimetric research might be most useful.

\subsection{Correlation between polarization and reflectance spectra \label{sec:dis1}}

Electromagnetic waves have three fundamental properties: intensity, frequency, and polarization. As each of them has a different sensitivity to a specific characteristic of scatterers, combining the information from more than one probe provides excellent complementary information for constraining the scattering environment \citep{Bohren1983}.  The polarization spectra $P_{\rm r}(\lambda)$ of large C-complex asteroids (Fig. \ref{Fig03}) already reveal some heterogeneity within a limited sample, motivating us to explore their correlation with reflectance spectra $I(\lambda)$.

To search for the correlation between $P_{\rm r}(\lambda)$ and $I(\lambda)$ for our target asteroids, we employed archival reflectance spectra from the second phase of the Small Main-belt Asteroid Spectroscopic Survey (SMASSII; \citealt{Bus2002,DeMeo2009}) and the Small Solar System Objects Spectroscopic Survey (S$^{\rm 3}$OS$^{\rm 2}$; \citealt{Lazzaro2004}). Figure \ref{Figap03} displays $I(\lambda)$ of the asteroids at the visible wavelength ($\sim$450--950 nm). The archives provide normalized reflectance, which is an observed spectrum divided by that of a solar analog, which is in turn normalized at 550 nm.
In theory, a reflectance spectrum may be constructed from our polarimetric observations; however, we did not obtain observations of solar analogs in a systematic manner: only a couple of solar analogs were observed in the course of the entire semester of observations and thus they do not necessarily have a similar airmass to the asteroids'. The mismatch in sky transparency can result in a discrepancy in the shape of the synthesized spectrum versus the existing reflectance data, thus we decided to use the archival data. Although a difference in phase angle between archival and our spectra may affect a slope correlation analysis, the phase-reddening effect (variation of the spectral slope with phase angle; \citealt{Gradie1986}) has been shown to be insignificant for asteroids analogous to our targets over a phase angle range of 0--30\degree\ \citep{Beck2021a}. We confirmed that for spectra available for multiple epochs for a single asteroid (Fig. \ref{Figap03}), the spectral shape is consistent within the measurement error, except for (145) Adeona, whose discrepancy seems systematic rather than phase-related. We thus assumed that phase-reddening effects on the asteroids in our sample are negligible and used average data points. Absolute polarization degree |$P_{\rm r}(\lambda)$| and reflectance $I(\lambda)$ were interpolated to match spectral resolution (450 to 900 nm with a 5 nm step) and normalized at 550 nm. To investigate the variation of |$P_{\rm r}(\lambda)$| with a change in the shape of $I(\lambda)$, we weight-averaged the |$P_{\rm r}(\lambda)$| values in each spectral bin (450--550 nm, 550--650 nm, 650--750 nm, and 750--900 nm) and measured a spectral slope $S$ in units of \% (100 nm)$^{\rm -1}$ in a given range.

\begin{figure}[!b]
\centering
\includegraphics[width=8.5cm]{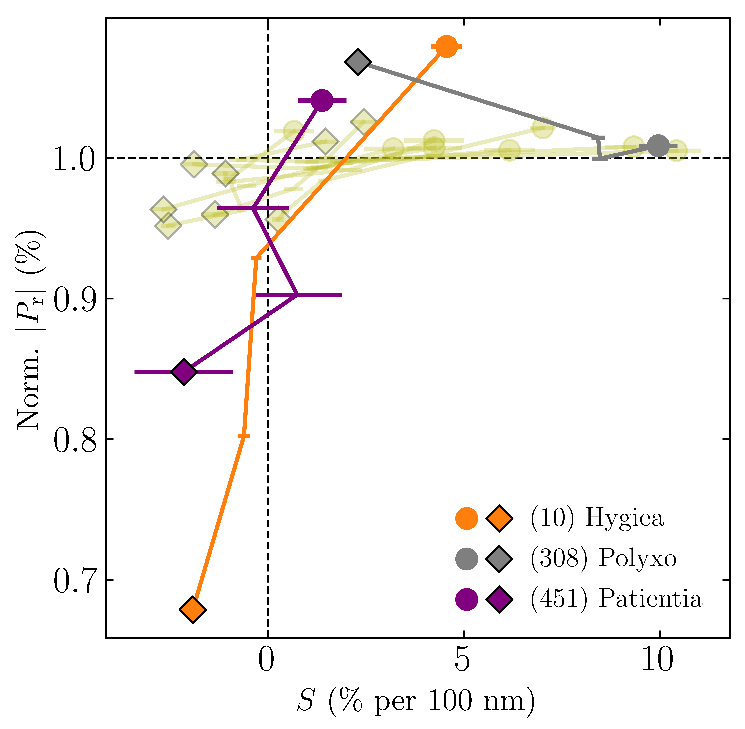}
\caption{Correlation between the spectra of the normalized absolute polarization degree |$P_{\rm r}(\lambda)$| and a spectral slope $S$ of reflectance spectra for the C-complex asteroids in Figures \ref{Fig03}. Both parameters were normalized at 550 nm and measured in each spectral bin. Circles and diamonds of each curve indicate the values derived in the shortest (450--550 nm) and longest (750--900 nm) bins, respectively. Including those endpoints, a single line connecting a total of four points represent |$P_{\rm r}$| and $S$ values of an asteroid evaluated over 450--550 nm, 550--650 nm, 650--750 nm, and 750--900 nm.
Only three asteroids -- (10) Hygiea, (308) Polyxo, and (451) Patientia -- are highlighted due to their significance while the rest with variations less than $\pm$5 \%\ of the unity (normalized |$P_{\rm r}(\lambda)$| at 550 nm) are not specified.
}
\label{Fig06}
\end{figure}

The distribution of the large C-complex asteroids in the |$P_{\rm r}(\lambda)$|--$S$ space is illustrated in Figure \ref{Fig06}. Circles and diamonds indicate the short- and long-ends of the spectral bins (450--550 nm and 750--900 nm, respectively). A single line for an asteroid connects four points, each of which represents the normalized |$P_{\rm r}$| and $S$ values with a 1$\sigma$ standard deviation per spectral bin.
At a glance, the majority of curves (colored in green) cluster within $\pm$5 \% of the horizontal line of unity, indicating a relatively neutral slope of their polarization spectra. Three asteroids showing appreciable variations are highlighted -- C-type (10) Hygiea, T-type (308) Polyxo, and C-type (451) Patientia -- and their ways of correlating |$P_{\rm r}$| and $S$ in the most part of the optical wavelength is twofold: the asteroids are located in either the first (upper right) or the third (lower left) quadrant in which |$P_{\rm r}(\lambda)$| and $S$ are positively correlated. In other words, lower (higher) |$P_{\rm r}$| values with regard to the normalization point at 550 nm are associated with negative (positive) spectral slopes. This is in contrast to siliceous S-complex asteroids with moderate-to-high albedos ($>$0.1) that obey the Umov law, thereby occupying the second and fourth quadrants \citep{Bagnulo2015,Bagnulo2023}.

\begin{figure}[!b]
\centering
\includegraphics[width=8.5cm]{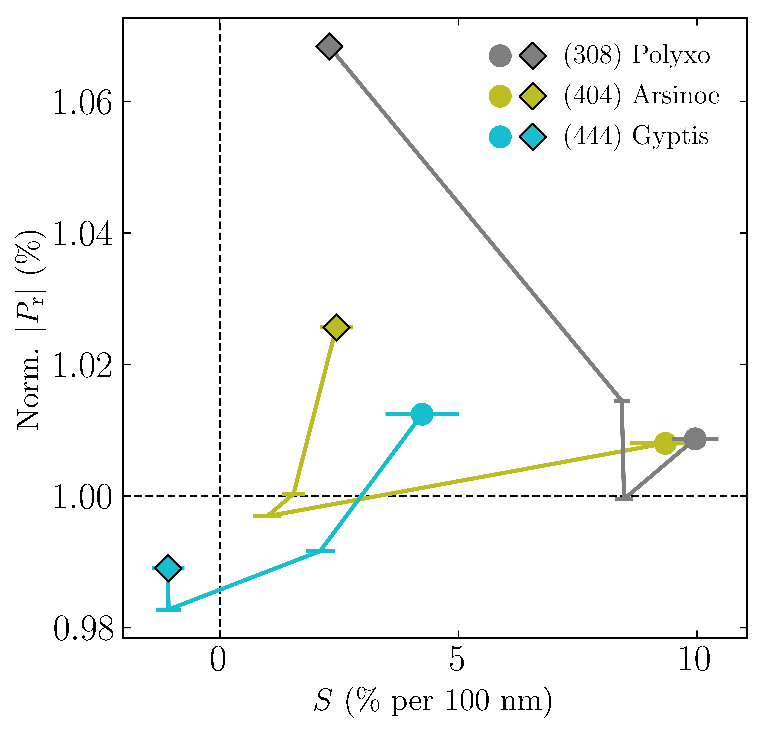}
\caption{Same as Fig. \ref{Fig06}, but for the asteroids showing convex-shaped polarization spectra in Fig. \ref{Fig03}.}
\label{Fig07}
\end{figure}

Interestingly, the asteroids showing dichotomic trends here correspond to the asteroids in a distinct shape of polarization spectrum (Fig. \ref{Fig03}): (10) Hygiea and (451) Patientia stand out for their steep blue polarization color, whilst (308) Polyxo has the largest concavity in its polarization spectrum. In Figure  \ref{Fig07}, we examine separately the dependence of asteroids with convex polarization spectra such as (308) Polyxo. Their stretch in parameter space is proportional to the concavity of the polarization spectrum shown in Figure \ref{Fig03}. Near 550--750 nm, (308) Polyxo and (404) Arsinoe exhibit a local V shape whose vertex appears around a 0.7-$\mu$m band center on their reflectance spectra. This indicates the presence of materials with different optical properties than those of the background components. (444) Gyptis, on the other hand, has the weakest concavity of the three and resembles the shape of other C-complex asteroids in Figure \ref{Fig06}.

The asteroids showing the dichotomy in the |$P_{\rm r}(\lambda)$|--$S$ correlation have the largest difference in $\alpha_{\rm 0}$ (either largest or smallest) values among the asteroids analyzed in Figure \ref{Fig06}. In Figure \ref{Fig05}a, the smaller-$\alpha_{\rm 0}$ part is predominantly occupied by F- and B-type asteroids with linear, featureless reflectance spectra in a negative-to-neutral spectral slope \citep{Gradie1982,Bus2002}, while the larger-$\alpha_{\rm 0}$ part is mostly occupied by Ch- and T-type asteroids having characteristic absorption bands in the 0.7-$\mu$m region in a neutral-to-moderately-positive color \citep{Rivkin2015b}. The finding that asteroids with a large difference in $\alpha_{\rm 0}$ also exhibit noticeable differences in the |$P_{\rm r}(\lambda)$|--$S$ space may suggest the presence of a common mechanism underlying the optical phenomena of large C-complex asteroids. 

\subsection{Linking polarimetric observations to surface properties of asteroids \label{sec:dis2}}

Multiple factors contribute to the optical properties of light scattered by asteroid regoliths. 
Research has long been conducted on how to enhance the NPB of atmosphereless small bodies. Laboratory experiments have shown that for low-albedo C-type analogs, a mixture of constituents in high albedo contrast develops a more pronounced depth and width of the NPB \citep{Shkuratov1987,Shkuratov1992,Shkuratov1994}. \citet{Shkuratov1987} introduced the partitioning factor $k$ = $\tilde{\omega_{\rm b}}$/$\tilde{\omega_{\rm d}}$ in his two-component mixture model as a means of assessing the albedo contrast between bright ($\tilde{\omega_{\rm b}}$) and dark ($\tilde{\omega_{\rm d}}$) materials. The author showed that a small-scale (the order of optical wavelengths) optical inhomogeneity (i.e., high $k$) of scatterers strongly amplifies the NPB in comparison to a single-component material. Recent laboratory studies (e.g. \citealt{Spadaccia2022}) validated the impact of this mixing effect on shaping the NPB.
$\alpha_{\rm 0}$ is also sensitive to the particle size of surface regoliths. The small (micron-to-submicron)-scale fine texture can affect the NPB parameters \citep{Dollfus1975,Dollfus1977} and particularly the NPB shows significant development in the presence of sub-micrometer grains on the surface \citep{Geake1990}; otherwise, the NPB is generally invariant to dust aggregates with a size of up to centimeters (\citealt{Gaffey1974,Spadaccia2022} and references therein).

The apparent concordance between the relative intensity of the NPB (Fig. \ref{Fig05}a) and the degree of aqueous alteration among different subclasses of C-complex asteroids \citep{Fornasier2014,Kwon2022b} motivates us to interpret the continuous difference in polarimetric properties in line with the progressive alteration induced by hydration of asteroids \citep{Rubin2007}. In Appendix \ref{sec:app5}, we compared laboratory measurements on anhydrous and hydrous regolith analogs and found supportive evidence that the hydrated samples tend to have larger $\alpha_{\rm 0}$ than the anhydrous ones over a wide range of size distributions.

Aqueous alteration is a low-temperature process \citep{Grimm1989} in which melting ice interacts with native anhydrous silicates (olivine and pyroxene), forming phyllosilicates (mainly saponite or serpentine depending on heating duration and initial water-to-rock ratio; \citealt{Zoensky1989,Kurokawa2020}). 
Infrared spectroscopy leverages the center wavelength and shape of a 3-$\mu$m band to diagnose the hydration status of asteroids. A sharp increase in a continuum over 2.8--3.0 $\mu$m implying the band centered at $<$3 $\mu$m indicates significant amounts of phyllosilicates on the surface regoliths, whereas a round-shaped band centered at $>$3 $\mu$m indicate negligible and/or lack of aqueous alteration of the regoliths (\citealt{Rivkin2022} and references therein). The former and the latter groups are referred to as `Sharp type (ST)' and `Non-sharp type (NST)', respectively, and the label for our C-complex asteroids is given in Table \ref{Fig03}. 
For asteroids devoid of pronounced alteration evidence (such as B and F types), their groundmass composition has been supposed to be similar to anhydrous carbonaceous chondrites whose optical properties are predominated by opaque carbon grains \citep{Gaffey1974,Gaffey1978,Rubin2007}.

\begin{figure}[!b]
\centering
\includegraphics[width=8.5cm]{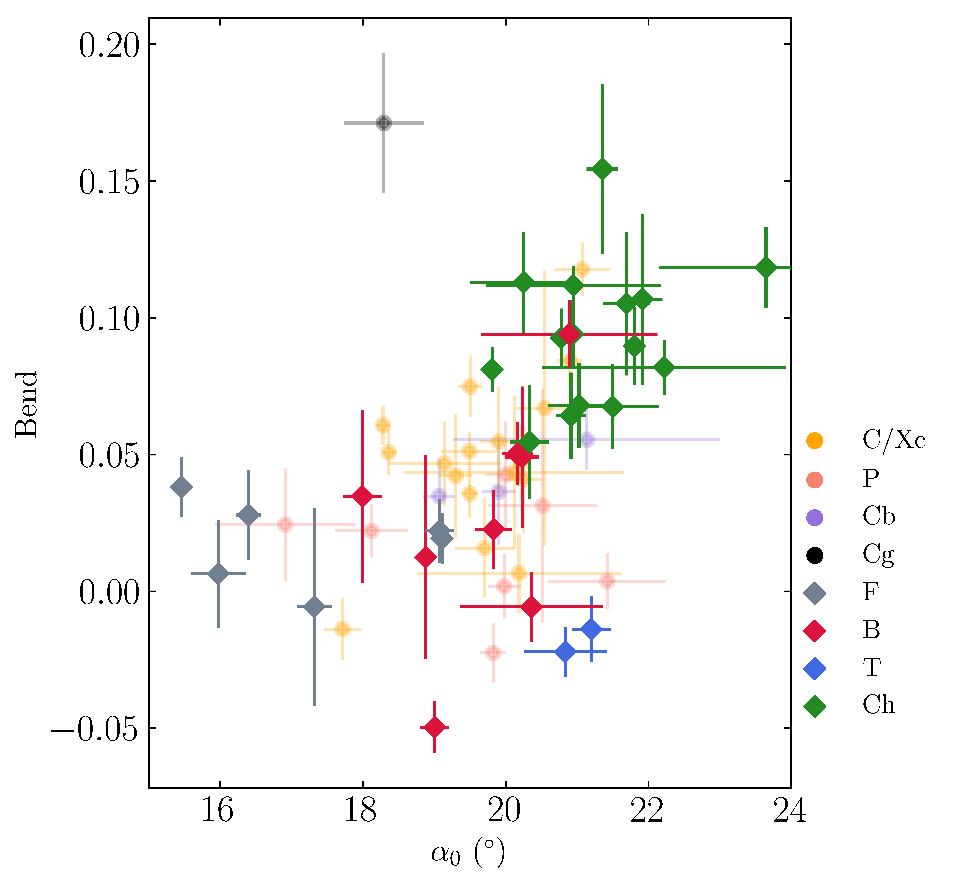}
\caption{Distribution of the modified Bend parameter (from \citealt{Gaffey1974,McCord1975a,McCord1975b}) as a function of $\alpha_{\rm 0}$ of the C-complex asteroids in Figure \ref{Fig05}. The two parameters are provided in Table \ref{t3}. }
\label{Fig08}
\end{figure}

The shape of a reflectance spectrum provides insight into the optical dominance of spectrally opaque materials which may also be related to the degree of hydration of the surface regolith.
\citet{McCord1975a,McCord1975b} introduced the `Bend parameter' as a measure of the curvature of the short visible domains (i.e., bending near 500 nm). An asteroid having a lower Bend parameter indicates higher opaques content, primarily carbon for C-complex asteroids. We measured the Bend parameter of the large C-complex asteroids whose archival reflectance spectra are available (Fig. \ref{Figap03}). While the authors in the original studies used spectra ranging from 320 nm to 1100 nm, most asteroids in modern archives have not been observed below $\sim$420 nm. We thus modified the equation to
\begin{equation}
{\rm Bend} \equiv (R_{\rm 560} - R_{\rm 440}) - (R_{\rm 730} - R_{\rm 560})
\label{eq:eq5}
\end{equation}
\noindent where $R_{\rm X}$ is the observed reflectance at the wavelength of X in nanometers. The second term in the first bracket on the right side was initially $R_{\rm 400}$ in \citet{McCord1975a,McCord1975b} but replaced here with $R_{\rm 440}$ to mitigate the extrapolation uncertainty. As a result, the modification would reduce the contrast in curvature when compared to reality if an absorption feature exists in the UV region. The calculated Bend parameters are tabulated in Table \ref{t3} and Figure \ref{Fig08} shows their distribution as a function of $\alpha_{\rm 0}$. Bend parameters are broadly distributed along the diagonal line from the lower left to the upper right. The average Bend parameters are 0.03$\pm$0.01, 0.02$\pm$0.05, and 0.09$\pm$0.02 for F-, B-, and Ch-type asteroids, respectively, whose increasing order is equivalent to the $\alpha_{\rm 0}$ distribution in Figure \ref{Fig05}a. The background objects (asteroids in Fig. \ref{Fig05}b) also align with the global trend. 

The observed subclass distribution in the $\alpha_{\rm 0}$--Bend parameter space would not be surprising if the surface composition is the primary factor. 
For F types having smaller $\alpha_{\rm 0}$ values, the absence of a sharp decrease in reflectance (UV dropoff) down to 0.35 $\mu$m is due to its opaque-rich assemblage that makes the UV absorption of Fe$^{2+}$ ions transition in silicates optically unrecognizable, yielding a small Bend parameter \citep{Burns1970,Johnson1973,Gaffey1978,Belskaya2005}. As regoliths undergo aqueous alteration (corresponding to a shift from F/B to Ch), $\alpha_{\rm 0}$ increases while such opaque minerals decline in optical dominance. UV dropoff becomes more robust and its wing extends to visible wavelengths, thereby producing a large Bend parameter.
The above scenario can make $\alpha_{\rm 0}$ and Bend parameters positively correlated with each other.

An alternative approach is to consider grain size effects. It is possible to qualitatively reproduce the observed Bend parameter distribution of different subclasses in Figure \ref{Fig08} if sizes of regolith constituents increase from Rayleigh-like sub-micrometer grains for F/B types (muting curvature) to $\sim$a few ten micrometer-sized grains for Ch types. However, the Bend and polarimetric parameters of the asteroids considered display no correlation with the albedo that is expected when grain size (and ensuing porosity) is the primary cause of diversity \citep{Johnson1973,Emery2004,Sultana2023}, let alone any trends in the intensity of silicate emission features around 10 $\mu$m \citep{Vernazza2012,Martin2022}. Neither can the size difference explain the dichotomy in the correlation between polarization and reflectance spectra (Fig. \ref{Fig06}). Therefore, although it is beyond the scope of this paper to rigorously evaluate the impact of particle size on the observed trends, we prefer the scenario that the regolith composition, particularly the difference in optical heterogeneity of asteroid regoliths induced by aqueous alteration, would be the primary cause of the observed polarimetric and spectroscopic properties. For a complete description of scattering processes by asteroid surfaces, both composition and size aspects must be taken into account.

As a side note, the T-type asteroids in Figure \ref{Fig08} have $\alpha_{\rm 0}$ comparable to those of Ch-type asteroids yet have Bend parameters as low as those of F/B-type asteroids. This would reflect the unique composition of the T-type asteroids (i.e., highly aqueously altered with abundant opaque minerals; \citealt{Kwon2022b} and references therein). 
It further suggests that $\alpha_{\rm 0}$ of large C-complex asteroids may be a proxy for optical heterogeneity in proportion to the degree of aqueous alteration in the regolith, but not necessarily for the total amount of opaque materials present.

\subsection{Potential of polarimetry as a tool for asteroid surface study \label{sec:dis3}}

It is impressive to observe that despite the narrow consideration of size and taxonomic classifications of asteroids in this study, their polarimetric dependence on phase angle and wavelength and their relationship with reflectance spectra demonstrate an appreciable degree of heterogeneity in their surface environments.
We should emphasize that the polarimetric parameters of an asteroid's regoliths are neither a simple nor a single-valued function for surface composition and the observational data are not yet sufficient to pinpoint a unique cause for the heterogeneity in the polarimetric behavior of large C-complex asteroids in the NPB. 
Nevertheless, the diversity of observations among the asteroids in this study suggests that the polarimetric properties of large C-complex asteroids in the NPB at optical wavelengths are primarily determined by their surface composition, which is strongly correlated with their hydration status. 

Our tentative conclusion may also explain the $\alpha_{\rm 0}$--diameter relationship seen in Figure \ref{Fig05}c. According to the existing $\alpha_{\rm 0}$--particle size relationship (e.g. \citealt{Geake1986,Geake1990}), the smaller and larger $\alpha_{\rm 0}$ values of F- and Ch-type asteroids are attributed to their bouldery ($>$millimeter) and fine-dust ($\lesssim$sub-micrometer) regolith particles, respectively, which appears to contradict previous studies.
However, if we look at size from the perspective of a dimension with a similar absorptivity (i.e., an imaginary part of the complex refractive index) on a wavelength scale, Ch-type regoliths can be viewed as clusters of small grains with high-albedo materials acting as a transparent matrix, whereas carbon-dominated F/B-type regoliths can be viewed as a single carbon grain. 
Light incident upon regoliths of the former and latter types will experience different numbers of interactions on a given scale, mimicking particles of small and large sizes, respectively.
This is only an educated guess at this point. A coordinated effort between laboratory and observational polarimetry studies is needed to understand better the mechanism behind light scattering on the asteroid surfaces.

Polarization parameters studied in our analysis seem to be applicable to smaller bodies, whose regolith size is expected to differ greatly from that of our target asteroids \citep{Delbo2007,Marchis2012,Gundlach2013}. (101955) Bennu is a sub-kilometer-sized Near-Earth Asteroid (NEA) and the target of NASA's sample return mission OSIRIS-REx \citep{Lauretta2017}. The NEA has small $\alpha_{\rm 0}$ and $|P_{\rm min}|$ best-fit values (17\fdg88$\pm$0\fdg40 and 1.43$\pm$0.14 \%, respectively), corresponding to F-type asteroids \citep{Cellino2018} and is apparently well located in the F-type regime in our Figure \ref{Fig05}a\footnote{(101955) Bennu has a B-type optical reflectance spectrum \citep{Lauretta2019,Hamilton2019} but an F-type polarization-phase curve \citep{Cellino2018}. Unlike the (barely) anhydrous large B/F-type asteroids examined in this study, whose dominant opaque material in the optical is carbon, this B/F-type NEA shows evidence of hydration comparable to CI chondrites \citep{Lauretta2019}, and magnetites are considered a key opaque, possibly existing in nanophase \citep{Trang2021}, that darkens and blues the optical spectra of the asteroid \citep{Hamilton2019}. The absence of an 0.7-$\mu$m hydration feature on Bennu indicates a higher temperature \citep{Hiroi1993,Hiroi1996} to which the asteroid was exposed during separation from its parent body. Hence, we present only the apparent agreement of Bennu’s NPB parameters in the context of the small $\alpha_{\rm 0}$--B/F taxonomy relationship proposed for large C-complex asteroids from this study yet refrain from further interpretation.}. 
\citet{Belskaya2005} showed that 38-km F-type (302) Clarrissa has a small $\alpha_{\rm 0}$ that is within the range of most of the F-type asteroids. Another NEA, (3200) Phaethon, the target of JAXA's DESTINY+ mission \citep{Arai2018}, is a $\sim$5.1-km \citep{Hanus2016} B-type asteroid having $\alpha_{\rm 0}$ of 20\fdg2$\pm$0\fdg2 and $|P_{\rm min}|$ of 1.2$\pm$0.1 \% \citep{Devogele2020}, also marginally residing in the B-type regime in Figure \ref{Fig05}a. 

Using $\alpha_{\rm 0}$ as a proxy to evaluate the relative primitiveness of an asteroid's surface appears consistent with a possible relationship between F-type asteroids and primitive surface environments \citep{Kolokolova1997,Cellino2018}.
133P/Elst-Pizarro is a $\sim$4-km \citep{Hsieh2009,Yu2020} main-belt comet, displaying comet-like dust ejection activity but dynamically in the asteroidal orbit \citep{Hsieh2006}. Its NPB shape resembles that of F-type asteroids \citep{Bagnulo2010} and likely the nucleus of 67P/Chyryumov-Gerasimenko \citep{Stinson2016}, casting the intriguing possibility of the association of F-type asteroids and a primitive comet nucleus whose surface is covered by an extremely low-tensile strength material \citep{Attree2018} with water-ice chunks \citep{Ciarniello2022,Fornasier2023}. A recent mid-infrared study of F-type (704) Interamnia with the smallest $\alpha_{\rm 0}$ in Figure \ref{Fig05} estimates volume fraction of water ice of 9--66 \% \citep{Haoxuan2023} supporting its primitive nature. 

Last but not least, the combination of the Bend parameter and $\alpha_{\rm 0}$ would help rough classification of F and B types. The absence of the UV dropoff information from the current archival data (SMASSII and S$^{\rm 3}$OS$^{\rm 2}$) makes F- and B-type asteroids hardly distinguishable from each other \citep{Mahlke2022}. Although the two subclasses are not sharply separated in Figure \ref{Fig08}, combining the two parameters is still able to isolate a moderate fraction of members of the two groups, hence encouraging further observations to increase the number of data sets available.

\section{Summary \label{sec:sum}}

This paper reports the optical (400--950 nm) spectropolarimetric properties of 12 large C-complex asteroids. In conjunction with archival polarimetric and spectroscopic data, we investigate the polarimetric behaviors around the backscattering region and their correlation with the spectroscopic properties of a total of 64 asteroids. The main results are as follows.

\begin{enumerate}
\item[1.] The normalized polarization spectra |$P_{\rm r}(\lambda)$| of 12 large C-complex asteroids observed with VLT/FORS2 and WHT/ISIS display color ranging from gray (0.01$\pm$0.01 \% per 100 nm) to blue ($-$0.12$\pm$0.01 \% per 100 nm) with a (quasi-)linear slope shortward of around 700 nm. The linearity extends to around 900 nm for most of the asteroids, except for three asteroids -- T-type (308) Polyxo, Ch-type (404) Arsinoe, and C-type (444) Gyptis which contain absorption bands in the so-called 0.7-$\mu$m and 3-$\mu$m region. The latter three have a concave |$P_{\rm r}(\lambda)$| shape with a local polarization minimum between 600 and 750 nm, corresponding to their 0.7-$\mu$m band center position.

\item[2.] The overall V-band polarization-phase curves $P_{\rm r}(\alpha)$ of our target asteroids in the NPB are broadly consistent, indicating the similarity of the bulk surface properties of the C-complex asteroids.

\item[3.] Despite the bulk similarity, there is an appreciable trend that different subclasses of the C-complex are distributed locally in the space of inversion angle and minimum polarization ($\alpha_{\rm 0}$--$|P_{\rm min}|$). The continuous alignment of F $\rightarrow$ B $\rightarrow$ T $\rightarrow$ Ch types in a direction of increasing both parameters ($\rho$ = 0.70$^{+0.06}_{-0.07}$ and $p$ = 0.0001) is particularly noteworthy since the ascending order reconciles the degree of aqueous alteration assessed by previous optical and infrared spectroscopic studies.
C/Xc-type asteroids fill the entire parameter space. 

\item[4.] Asteroids in the particulate domain display weak, but meaningful inverse correlations with their diameters ($\rho$ = $-$0.42$^{+0.08}_{-0.07}$ and $p$ = 0.01), primarily driven by the distribution of C/Xc-type asteroids ($\rho$ = $-$0.51$^{+0.13}_{-0.11}$ and $p$ = 0.04). A larger number of asteroids is needed to confirm this trend.

\item[5.] Asteroids' albedos show no statistically significant relationship with any other parameters (including the NPB parameters, diameters, or orbital elements), in agreement with previous polarimetric studies.

\item[6.]  For asteroids with sufficient variations in polarization, we observe a dichotomous relationship between their polarization and reflectance spectra. The spectra of absolute polarization degree |$P_{\rm r}(\lambda)$| and a spectral slope $S$ are positively correlated, but in an opposite direction: either having higher |$P_{\rm r}(\lambda)$| and positive $S$ or having lower |$P_{\rm r}(\lambda)$| and negative $S$. The former group, represented by (308) Polyxo, has a concave polarization spectrum and relatively large $\alpha_{\rm 0}$ values, whereas the latter group has a linear polarization spectrum and smaller $\alpha_{\rm 0}$ values.
None of the asteroids exhibit an inverse correlation between polarization and reflectance spectra like S-type asteroids \citep{Bagnulo2015,Bagnulo2023} that obey the Umov law.


\item[7.] The relationship of the Bend parameter (modified in this study from the one originally defined by \citealt{McCord1975a,McCord1975b}) and $\alpha_{\rm 0}$ display a broad proportionality for the asteroids in our sample. Given that the Bend parameter evaluates from the curvature around 500 nm the relative influence of opaque materials on reflectance spectra, this trend indicates the sequence of F $\rightarrow$ B $\rightarrow$ T $\rightarrow$ Ch-type asteroids in $\alpha_{\rm 0}$ would reasonably well respond to their petrologic difference.

\end{enumerate}

There is still a significant lack of information on the polarization distribution of asteroids covering a wide range of wavelengths, compared to other photometric or spectral archival data. 
For this reason, it is essential to put continuous efforts into building polarimetric catalogs (e.g. \citealt{Bendjoya2022,Lupishko2022,Masiero2022}) in order to refine the interpretation of scattered light and eventually to define working observables for asteroid surfaces. 
\\

\begin{acknowledgements}

We thank an anonymous referee for providing valuable suggestions that led to substantial improvements. Discussions with Johannes Markkanen regarding light scattering in laboratory analogs were useful to elaborate the manuscript.
Y.G.K gratefully acknowledges funding from the Volkswagen Foundation. Based in part on observations made with the ESO Very Large Telescope at the Paranal Observatory under program 095.C-0925 and 097.C-0853 (PI: Bagnulo) and collected at the WHT operated on the island of La Palma by the Isaac Newton Group during semester 14A.
\end{acknowledgements}


\begin{appendix}

\section{Selection of taxonomic types of asteroids in Table \ref{t3} \label{sec:app1}}
\counterwithin{figure}{section}

The purpose of this section is to demonstrate our criteria for selecting the taxonomic types of asteroids. We referred to the taxonomic types of the two classification systems in the first place from the table provided by Small-Body Database Lookup\footnote{\url{https://ssd.jpl.nasa.gov/tools/sbdb_lookup.html#/}} and then applied following criteria to re-classify asteroids that have been designated as different subclasses by Tholen \citep{Tholen1984} and Bus-DeMeo (SMASSII; \citealt{DeMeo2009}) classification systems.

Firstly, since the contemporary SMASSII catalog does not cover the UV--short visible region, in which Tholen F types exhibit a different spectral slope from SMASSII B types, Tholen F types were grouped with SMASSII B types.
We thus always prioritized Tholen F types over SMASSII B types unless both classification systems assign B to an asteroid (e.g. (372) Palma). 

Secondly, SMASSII X types contain a higher heterogeneity of member asteroids, as the taxonomic scheme uses only spectral shapes and does not incorporate albedo information. Tholen P types fall in this spectral group. SMASSII X types' spectral slopes distribute over a wide range from a neutral-to-slight-red (C-type) slope to a moderately red (Tholen P-type) slope \citep{DeMeo2009}. Therefore, for asteroids assigned in Tholen P types and/or SMASSII X, Xc, or Xk types, we re-classified them based on the mean reflectance data available on the SMASSII webpage\footnote{\url{http://smass.mit.edu/smass.html}} and previous X-type spectroscopic observations by \citet{Fornasier2011}. If a spectral slope of an asteroid over 500--800 nm is $\gtrsim$2.5 \% per 100 nm, we assigned it P and otherwise Xc. Xc types were illustrated in Figures \ref{Fig05} and \ref{Fig08} using the same symbol as the C types, given that their spectral characteristics are similar to those of the C-type asteroids \citep{DeMeo2009,DeMeo2015}. C types contain features shared by several subclasses and were therefore ranked lowest. 

Lastly, the technical definition of SMASSII Ch types (and some of Tholen G types) is when an asteroid has an absorption band in the so-called 0.7-$\mu$m region associated with phyllosilicates. However, this feature has been observed not only in Ch-type asteroids but also in other taxonomic subclasses, and sometimes a single asteroid shows variations in its band depth at different epochs of observations (e.g. \citealt{Fornasier2016}). Given that this wavelength domain overlaps with the atmospheric H$_{\rm 2}$O absorption band \citep{Lord1992}, thereby subject to systematic uncertainties introduced by calibration of observations, the 0.7-$\mu$m band feature reported for some asteroids would be artifacts (e.g. \citealt{Vilas1989}). In this study, if an asteroid has ever been diagnosed as Ch by either of the two taxonomies, we assigned it Ch. However, for asteroids that have a 0.7-$\mu$m absorption feature but belong to non-Ch types, we determined their taxonomies considering infrared spectroscopic observations together: if such an asteroid has a V-shaped absorption feature whose band center is $<$3 $\mu$m (phyllosilicate-dominated `ST' in Table \ref{t3}), we took its 0.7-$\mu$m band as valid evidence for the hydrated status of the asteroid and thus designated it as a Ch type. On the contrary, if an asteroid has a 0.7-$\mu$m feature but not an ST-like infrared absorption (either having an infrared band center at $>$3 $\mu$m, i.e., `NST' in Table \ref{t3}, or no available infrared data), we kept its non-Ch spectral type. 

Asteroids whose taxonomic type was updated on the basis of the above criteria are footnoted with stars ($\star$) in Table \ref{t3}. Below is special remarks for three asteroids.
\begin{enumerate}
\item[$\bullet$]{\bf (54) Alexandra and (386) Siegena}~~~~~Both are Tholen C and SMASSII C types. Despite their non-Ch type designations, their reflectance spectra contain both a discernible 0.7-$\mu$m absorption feature \citep{Fornasier2014} and an ST-like phyllosilicate absorption band \citep{Takir2012,Rivkin2022}. We thus assigned them Ch.\\

\item[$\bullet$]{\bf (62) Erato}~~~~~This is Tholen BU and SMASSII Ch type. The asteroid is a member of the Themis asteroid family \citep{Nesvorny2015} and should be considered a Ch type since SMASSII taxonomy already designates the asteroid as Ch. However, given the absence of a 0.7-$\mu$m absorption feature and its overall affinity to the B-type reflectance spectra (\citealt{Yang2009} and references therein), we considered this asteroid as a B subclass.\\

\end{enumerate}

\section{Statistical estimation to test for the correlations in Figure \ref{Fig05} \label{sec:app2}}
\counterwithin{figure}{section}

To discern whether a set of two variables in Figures \ref{Fig05}a and \ref{Fig05}c are correlated, we conducted Spearman's rank correlation test \citep{Spearman1904} whose correlation coefficient $\rho$ and p-value $p$ were estimated using Monte Carlo simulation. Given a data set consisting of $N$ data pairs, $X$ $\pm$ $\sigma_X$ and $Y$ $\pm$ $\sigma_Y$, each $x_{\rm i}$ and $y_{\rm i}$ pair was randomly generated in the Gaussian distribution whose standard deviation equals to the 1-$\sigma$ uncertainty, $\sigma_X$ and $\sigma_Y$, of the central values. For statistical significance, we tested $i$ = 10,000 Monte Carlo particles creating the probability distribution of $\rho$ and $p$. 

\begin{figure}[!b]
\centering
\includegraphics[width=8.5cm]{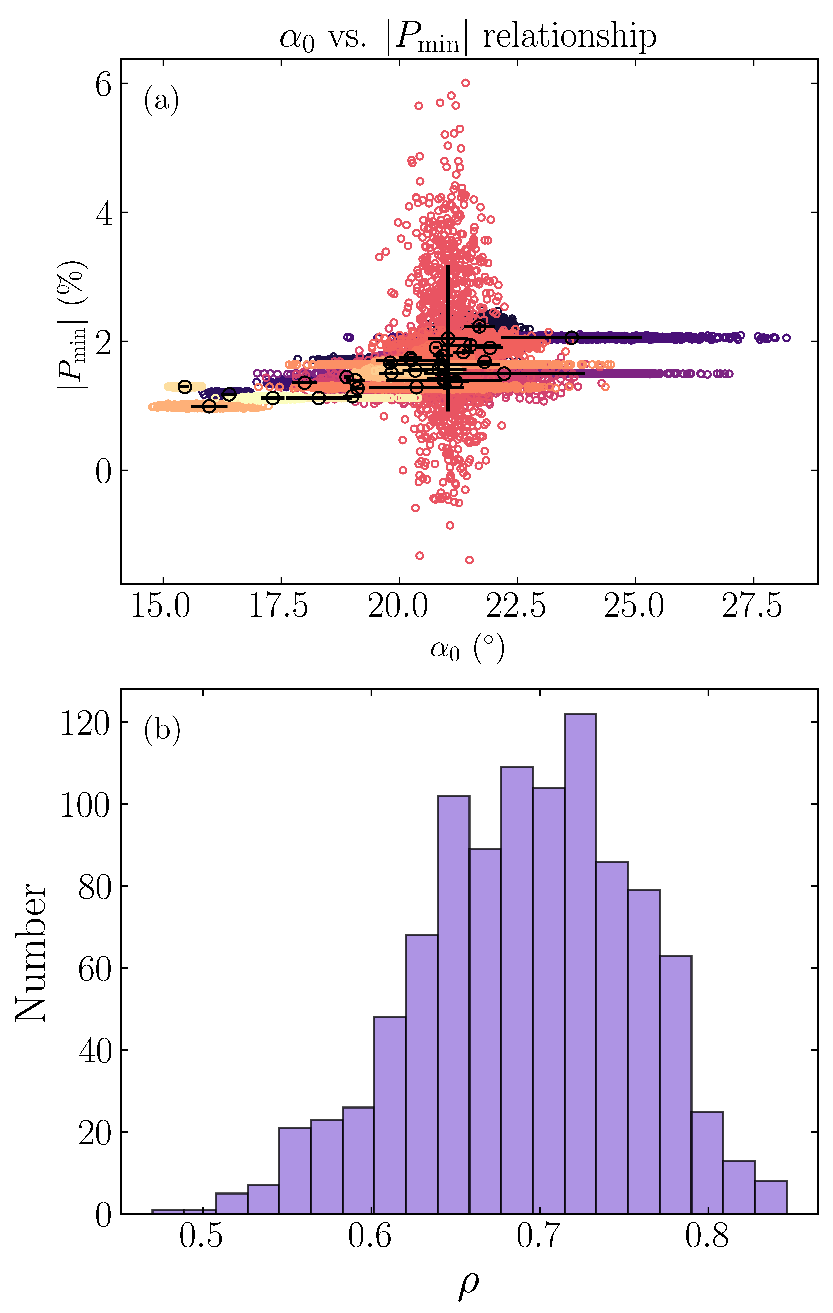}
\caption{Number distribution of 10,000 clones randomly generated within the range corresponding to the errorbars of the inversion angle $\alpha_{\rm 0}$ and the absolute value of the minimum polarization degree |$P_{\rm min}$|. The retrieved Spearman's rank correlation coefficient $\rho$ histogram is present in panel b. Colors do not follow the color code used in Figure \ref{Fig05}a.
}
\label{Figap01}
\end{figure}

\begin{figure}[!b]
\centering
\includegraphics[width=8.5cm]{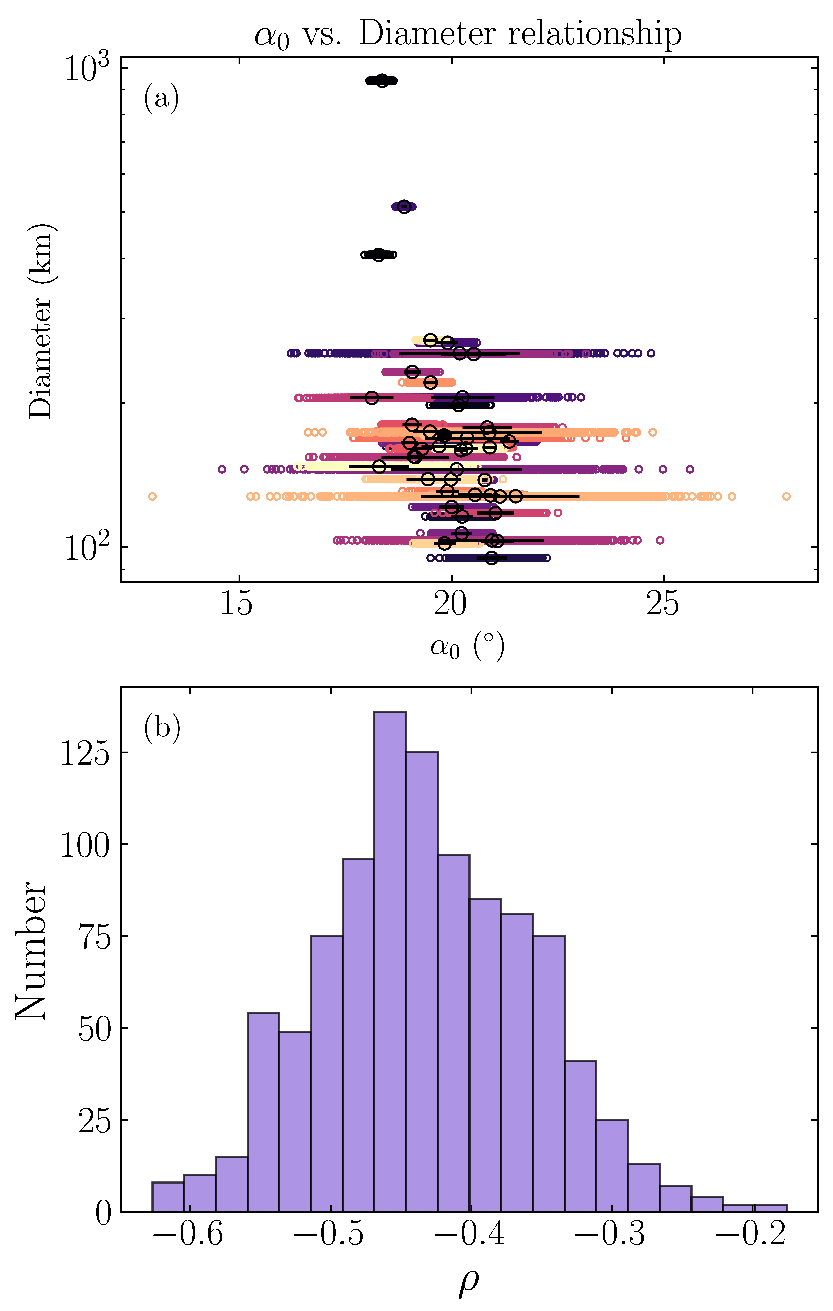}
\caption{Same as Figure \ref{Figap01} but the test for the correlation of $\alpha_{\rm 0}$--Diameter in Figure \ref{Fig05}c.
}
\label{Figap02}
\end{figure}

Figure \ref{Figap01} shows the distribution of 10,000 clones generated to test the correlation of $\alpha_{\rm 0}$--|$P_{\rm min}$| in Figure \ref{Fig05}a. In a 68.2 \% confidence interval, the median of the distribution is $\rho$ = 0.70$^{+0.06}_{-0.07}$ and its mode is 0.72. The resulting p-value is $p$ = 0.0001, confirming the proposed correlation's statistical significance ($>$3$\sigma$). Likewise, the identical scheme of the Monte Carlo method reveals that $\rho$ = 0.29$^{+0.07}_{-0.11}$ and its mode of 0.25 for F types, $\rho$ = 0.43$^{+0.25}_{-0.18}$ and its mode of 0.42 for B types, and $\rho$ = 0.36$^{+0.17}_{-0.18}$ and its mode of 0.33 for Ch types. T types were not analyzed due to the small data number ($N$ = 2). Similar to the trend of F--B--T--Ch-type asteroids as a whole, members of asteroids in each taxonomic type are in a positive correlation between the NPB parameters, but with less statistical significance. 

The correlation test between the inversion angle $\alpha_{\rm 0}$ and the diameter of asteroids in Figure \ref{Fig05}c is also shown in Figure \ref{Figap02}. For the asteroids whose $\alpha_{\rm 0}$ values lead them to belong in the particulate domain \citep{Geake1986}, the median of the distribution as a whole in a 68.2 \% confidence interval is $-$0.42$^{+0.08}_{-0.07}$ and its mode is $-$0.46, with a p-value of 0.01. This trend seems largely driven by the distribution of C-type asteroids ($\rho$ = $-$0.51$^{+0.13}_{-0.11}$ and its mode of $-$0.52, with a p-value of 0.04) distributed in the widest range of diameters. 

\section{Archival reflectance spectra of the asteroids in Table \ref{t3} \label{sec:app3}}
\counterwithin{figure}{section}

This section summarizes the reflectance spectra normalized at 550 nm used in the analysis in the main text. In Figure \ref{Figap03}, the spectra are given in ascending order of the number of asteroids.

\begin{figure*}[!htb]
\centering
\includegraphics[width=0.95\textwidth]{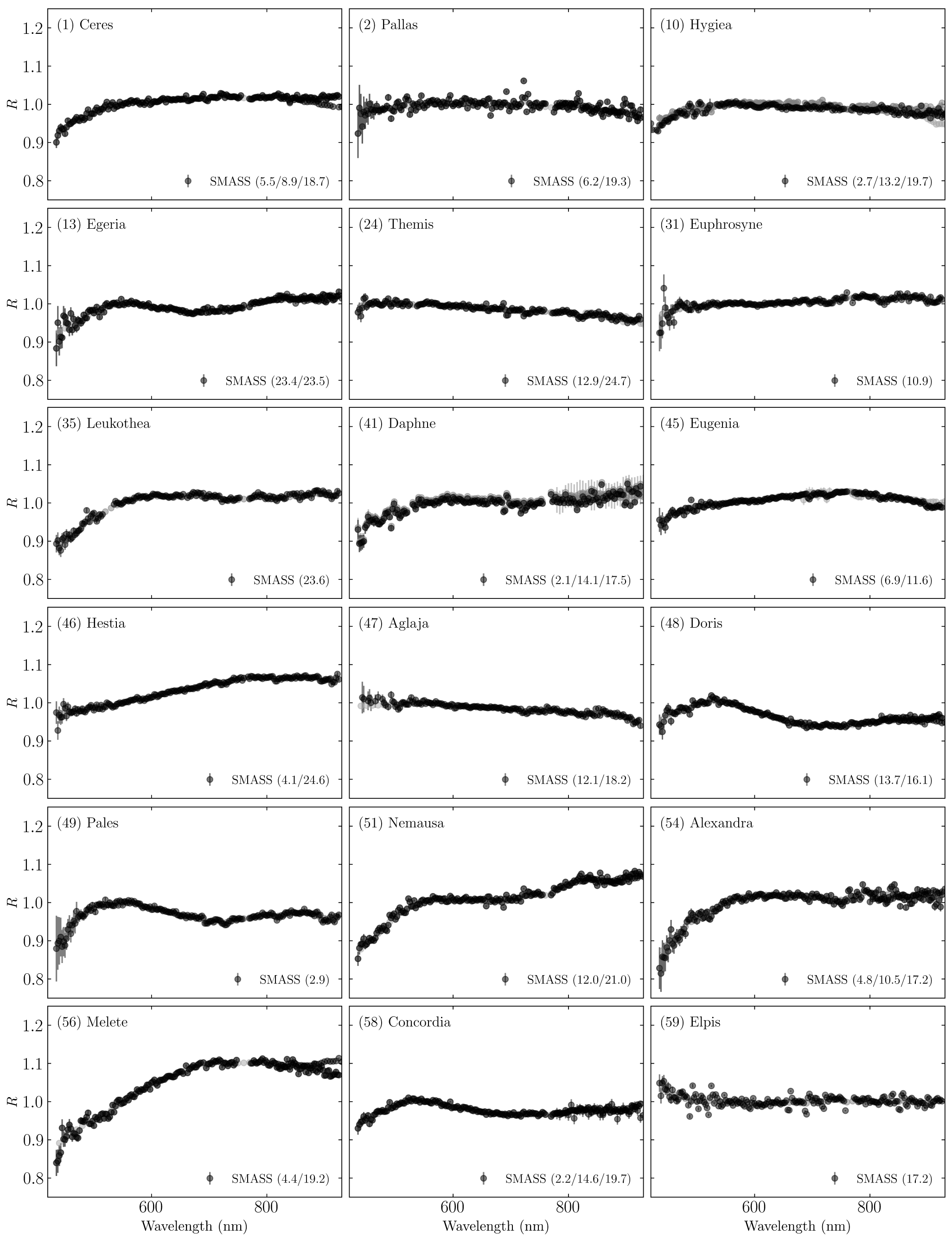}
\caption{Reflectance spectra of large, low-albedo C-complex asteroids in Table \ref{t3} normalized at 550 nm from SMASSII \citep{Bus2002,DeMeo2009} and S$^{\rm 3}$OS$^{\rm 2}$ \citep{Lazzaro2004}. Numbers in the parenthesis in each panel indicate phase angles at the time of observations. Weak features near 900 nm appear in the spectra taken at different times with different instruments at edge wavelengths where detectors are less sensitive. Discontinuities shown in several SMASSII spectra near 800 nm are caused by the concatenation of different datasets \citep{DeMeo2009}.}
\label{Figap03}
\end{figure*}

\addtocounter{figure}{-1}
\begin{figure*}[!htb]
\centering
\includegraphics[width=0.95\textwidth]{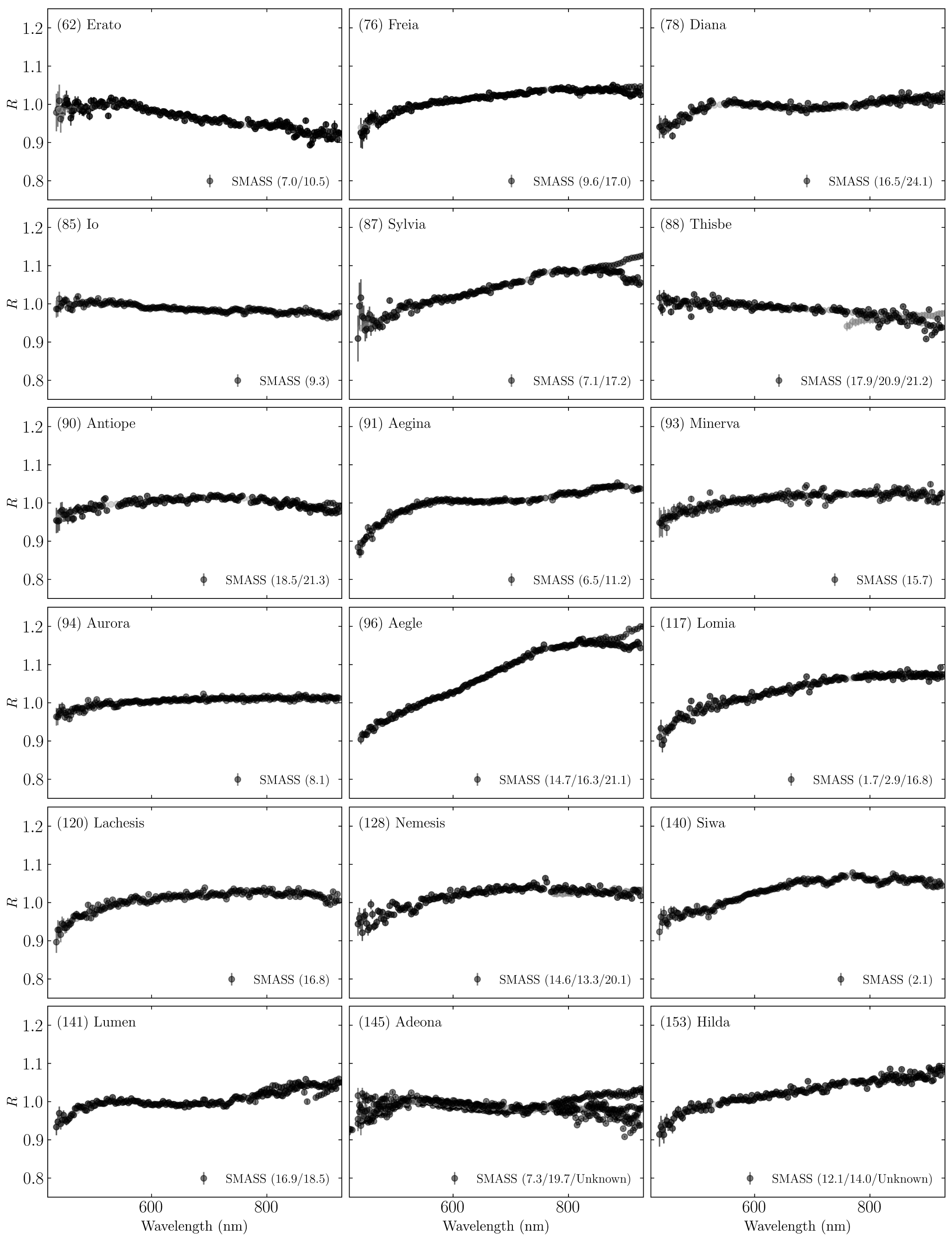}
\caption{Continued.}
\end{figure*}
\addtocounter{figure}{-1}
\begin{figure*}[!htb]
\centering
\includegraphics[width=0.95\textwidth]{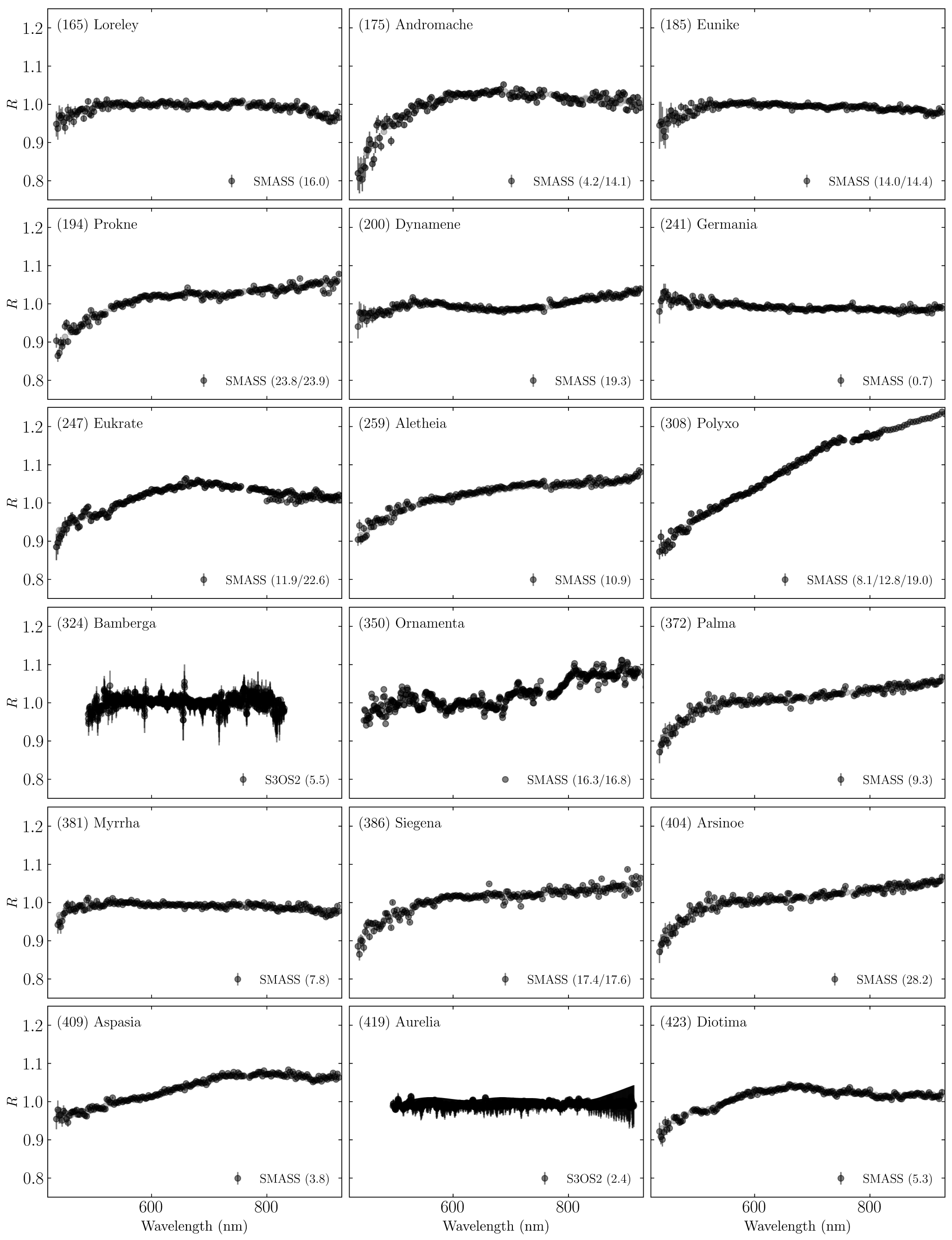}
\caption{Continued.}
\end{figure*}
\addtocounter{figure}{-1}
\begin{figure*}[!htb]
\centering
\includegraphics[width=0.95\textwidth]{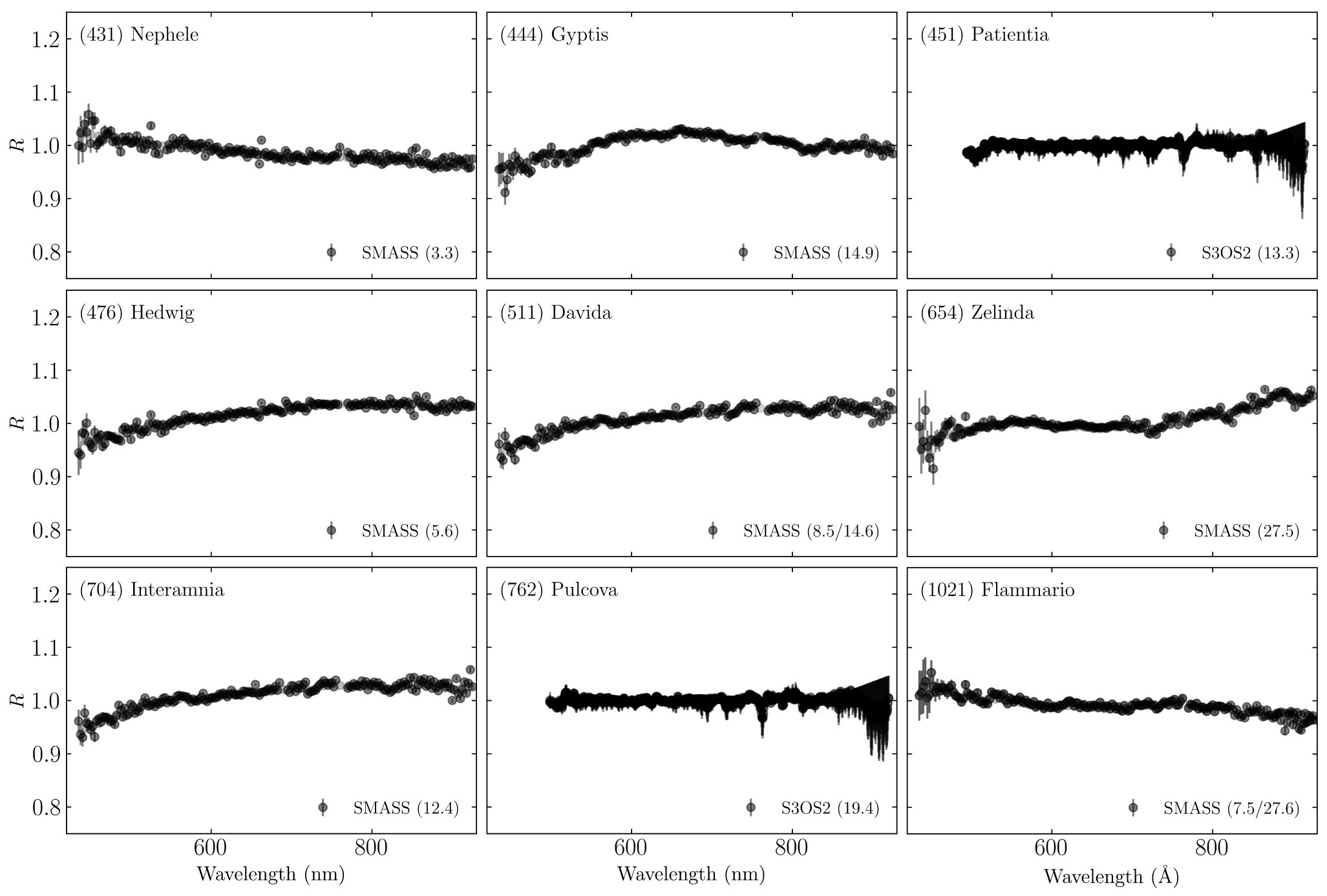}
\caption{Continued.}
\end{figure*}
\renewcommand{\thefigure}{\arabic{figure}}
\clearpage

\section{Comparison of the laboratory experiments on the NPB properties of anhydrous and hydrous regolith analogs} \label{sec:app5}
\counterwithin{figure}{section}

\begin{figure*}[!b]
\centering
\includegraphics[width=12cm]{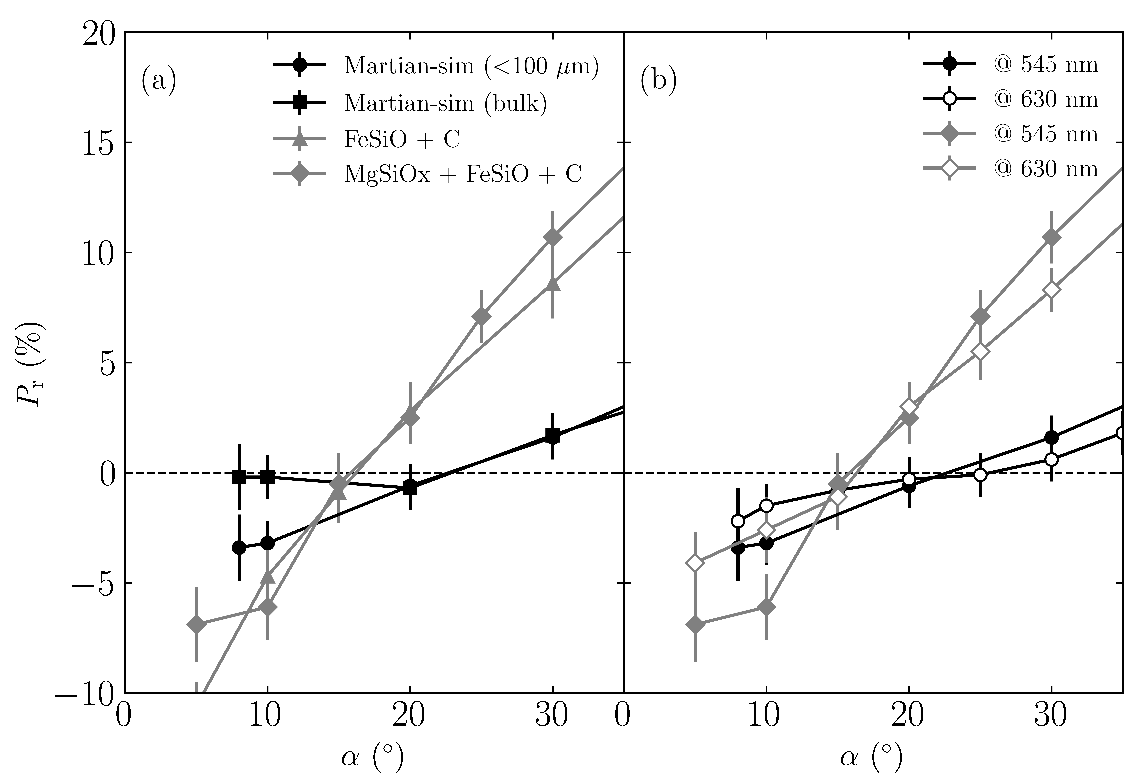}
\caption{Polarization-phase curves of martian simulants and mixtures of Mg-Fe silica with carbon from the PROGRA$^{2}$ experiment database. Sample names in the legend are the same as those in the database. Data in panel a are measurements at 545 nm. Filled symbols in panel b are the same as those in panel a (i.e., the Martian simulant ($<$100 $\mu$m) and Mg$\cdot$Fe-silica with carbon at 545 nm) but their open counterparts indicate measurements at 630 nm.}
\label{Figap04}
\end{figure*}

Laboratory measurements of samples with properties that are similar to those before and after aqueous alteration may provide additional information regarding the $\alpha_{\rm 0}$ development. As such, we utilized experiment data from PROGRA$^{2}$ (the French acronym for Optical Properties of Astronomical and Atmospheric Grains\footnote{\url{https://www.icare.univ-lille.fr/progra2/}}). The database provides polarization and brightness phase curves for irregular aggregates suspended or deposited on surfaces over 450--1600 nm and has been widely used to study light scattered by small solar system objects and Martian regoliths \citep{Worms2000,Hadamcik2006,Levasseur-Regourd2015,Hadamcik2023}. 

In Figure \ref{Figap04}a, we show the polarization-phase curves of two martian soil simulants (consisting of mainly silica and Fe oxides, with a size distribution of $<$100 $\mu$m and 250--1000 $\mu$m, each), Fe-silica with carbon (with a size distribution, peaked at $\sim$50 $\mu$m having a long tail up to $\sim$300 $\mu$m), and Mg$\cdot$Fe-silica (25 \% of each) with carbon (14 nm, 50 \% in abundance) with a size distribution peaked at $\lesssim$20 $\mu$m. The last two samples are analogs of F-type asteroids \citep{Gaffey1978,Allen1998,Hadamcik2003} in that their optical properties are dominated by intimately mixed opaques (nanometer-sized carbon) and thus exhibit decent bulk optical homogeneity. Their polarization-phase curves are highly similar despite the Mg-silica in the latter samples, which makes sense given that carbon is so optically dominant that even small amounts (a few percent) are sufficient to flatten and mute the spectrum and lower the albedo well under 0.1 \citep{Johnson1973,Gaffey1974,Mukai1986}. The Martian simulant (composed of olivines, pyroxene, and oxides) contains OH bands on its VNIR reflectance spectrum, albeit not exactly produced by phyllosilicates \citep{Allen1998,Hadamcik2003}. 
We thus expect the Martian simulant to exhibit greater optical heterogeneity at the microscale than the silica samples, qualitatively mimicking a composition post-aqueous alteration (higher optical heterogeneity of the constituents with hydrous bands), especially when compared to the other two F-type analogs.

Despite having significantly different size distributions, samples in comparable optical heterogeneity possess nearly identical $\alpha_{\rm 0}$ values (Fig. \ref{Figap04}). $\alpha_{\rm 0}$ values are consistently smaller for optically homogeneous, small-sized opaque-governing samples (grey lines) than for optically heterogeneous, oxide-containing samples (black lines), in agreement with recent laboratory experiments (e.g. \citealt{Spadaccia2022}). The polarization color in all samples in the NPB is blue over 545--630 nm (panel b in Fig. \ref{Figap04}), which is consistent with the observations of our target asteroids (Fig. \ref{Fig03}). 
The samples compared here may not be suitable for testing the impact of submicron-sized particles on the NPB shape due to the practical reasons for sample preparation (J. Markkanen, personal communication); nonetheless, the observed $\alpha_{\rm 0}$ differences between samples with varying degrees of carbon impact on the wavelength-scale support the hypothesis that $\alpha_{\rm 0}$ would tie to the compositional properties of C-complex asteroids.

\end{appendix}

\end{document}